# Surface potential-driven surface states in 3D topological photonic crystals

Haedong Park[1,2], Sang Soon Oh[2]*, and Seungwoo Lee[1,3,4,5]*

[1]KU-KIST Graduate School of Converging Science and Technology, Korea University, Seoul 02841, Republic of Korea

[2]School of Physics and Astronomy, Cardiff University, Cardiff CF24 3AA, United Kingdom

[3]Department of Biomicrosystem Technology, Korea University, Seoul 02841, Republic of Korea

[4]Department of Integrative Energy Engineering and KU Photonics Center, Korea University, Seoul 02841 Republic of Korea

[5]Center for Opto-Electronic Materials and Devices, Post-Silicon Semiconductor Institute, Korea Institute of Science and Technology (KIST), Seoul 02792, Republic of Korea

*Email: ohs2@cardiff.ac.uk, seungwoo@korea.ac.kr

**Abstract**: Surface potential in a topological matter could unprecedentedly localize the waves. However, this surface potential is yet to be exploited in topological photonic systems. Here, we demonstrate that photonic surface states can be induced and controlled by the surface potential in a dielectric double gyroid (DG) photonic crystal. The basis translation in a unit cell enables tuning of the surface potential, which in turn regulates the degree of wave localization. The gradual modulation of DG photonic crystals enables the generation of a pseudomagnetic field. Overall, this study shows the interplay between surface potential and pseudomagnetic field regarding the surface states. The physical consequences outlined herein not only widen the scope of surface states in 3D photonic crystals but also highlight the importance of surface treatments in a photonic system.

**Main text**:

The discovery of topological insulators opened exciting new realms of physics[1,2], and extensive efforts have been made to understand topological physics over the last decade. At this field's core, it is important to note zero-dimensional degeneracies such as Dirac[3-9] or Weyl points[10-19] and one-dimensional degeneracies such as nodal lines[20-28]. These band degeneracies have been theoretically or experimentally realized using metals[21,29], semimetals[30-36], phononic crystals[37-39], electrical circuits[40], and photonic crystals[10,11,18,41-49]. Based on the definition of topological insulators, many studies propose several boundary states, such as one-way surface/edge states[39,50,51], drumhead surface states[40,52-55], and Fermi arcs[47,48,56-59].

Beyond them, one can adjust the boundary state with surface/edge potential[60-67] and pseudomagnetic field[13,37,68]. In condensed matters, the surface/edge potential[60-64] arises from the passivation on the surface/edge of a given material[65-67]. The surface/edge potential[60-63] has been successfully implemented in two-dimensional (2D) materials including graphene[65], boron nitride[66], and semimetals;[60,64,67] consequently, the excitation and manipulation of surface/edge states came to the fore. However, such driving and using surface potential are yet to be exploited in topological photonic crystals.

Meanwhile, the pseudomagnetic field is a virtual effective field stemming from a spatial reconfiguration of the crystal lattice without a real magnetic field[69-72]. Like a real magnetic field enabling the quantum Hall effect[73,74], the pseudomagnetic field was found to be a gold vista for topologically nontrivial surface/edge states in semimetals[75-79], photonic crystals[80-82] and phononic crystals[37,70,71,83]. Nevertheless, the surface state's profiles along the zeroth

Landau level on a 3D Weyl photonic crystal, driven by the pseudomagnetic field, were out of reach.

Here, we demonstrate photonic surface waves that arise from the interplay between surface potential and pseudomagnetic field in photonic systems. First, we investigate the effect of the surface potential in a system governed by the Weyl equation with a pseudomagnetic field. The pseudomagnetic field is switched on by the Weyl points that vary with the unit cell positions along the boundary-boundary direction in the system. Then, we apply the effective Hamiltonian description to a photonic system based on double gyroid (DG) photonic crystals[17]. The DG photonic crystals are found to exhibit Weyl points due to the geometrical perturbation, which appears as a defect-like shape. To realize a pseudomagnetic field, we constitute a photonic array of the DG unit cells with a spatial gradient of the perturbation, i.e., the degree of the defect varies with the position. We then compute quantized Landau levels and eigenstates along the zeroth Landau level to quantitate the asymmetric localization of photonic waves on the surfaces. This observed asymmetry in wave localization evidences the existence of the surface potential. Then, we tune the translation of the basis of the unit cell to tame the surface termination. This tuning exquisitely controls the surface potential so that the degree of wave localization varies with respect to the tuning. Finally, we implement such results to the evasion behaviors of a photonic wave to observe the interplay between the surface potential and pseudomagnetic field.

**The attraction of waves by the surface potential in the Weyl system**

First, let us describe the pseudomagnetic field's effects using the Weyl Hamiltonian. For this, we consider a system periodic along the $x_1$- and $x_2$-directions for an orthogonal coordinate system. The system consists of $N$ unit cells along the $x_3$-direction (the horizontal rightward direction in Fig. 1a-b), and surface boundaries are parallel to the $x_1$- and $x_2$-directions. We assume that this system is governed by

$$\boldsymbol{\sigma} \cdot (-i\nabla - \mathbf{k}_w)\psi = i\sigma_0 \frac{\partial \psi}{\partial t} \tag{1}$$

that describes a Weyl point at $\mathbf{k}_w$, where $\sigma_0$ is the 2×2 identity matrix, and $\sigma_i$ ($i = 1, 2, 3$) are the Pauli matrices. We decompose the Weyl point's location $\mathbf{k}_w$ into $\mathbf{k}_w = \mathbf{k}_{w,0} + \mathbf{A}_w^n$. $\mathbf{k}_{w,0}$ is a constant while $\mathbf{A}_w^n$ depends on the unit cell's index $n$. We assume that $\mathbf{A}_w^n$ linearly varies with a specific value $p$ (henceforth, this value is referred to as perturbation strength) which also linearly varies with $n$:

$$\mathbf{A}_w^n \propto p(n) = p_s n + const, \tag{2}$$

where $p_s$ is a proportional constant (refer to Fig. 1a-b). Then, the pseudomagnetic field can be written as $\mathbf{B} = \nabla \times \mathbf{A}_w^n$, which is proportional to $p_s$.

Let us suppose two systems with different signs of slopes, i.e., one system described by $p_s = -p_{s,0}$ (Fig. 1a) and the other described by $p_s = +p_{s,0}$ (Fig. 1b) where $p_{s,0}$ is positive. Their $p(n)$ are schematically illustrated as the size of yellow circles in Fig. 1a-b. The pseudomagnetic fields $\mathbf{B}$ for these two are in the same magnitudes and in opposite directions. Their wave localization of the zeroth Landau level at a specific point in the momentum space shows mutually symmetric characteristics for these opposite fields, as shown in Fig. 1c.

Now, let us implement a surface potential $V$ as follows:

$$V = V_s \sigma_3, \tag{3}$$

where $V_s$ is nonzero only around the boundaries[60,61,64-66], like Fig. 1d. When we include the surface potential term with $V_s$, further detailed in Methods, eq. (1) is rewritten as:

$$\{\boldsymbol{\sigma} \cdot (-i\nabla - \mathbf{k}_w) + V\}\psi = i\sigma_0 \frac{\partial \psi}{\partial t}. \tag{4}$$

By applying the same $V$ for the two systems in Fig. 1a-b, we can observe mutually asymmetric localization, as shown in Fig. 1e. The two localized states commonly show a slight shift in the left direction, implying that the surface potential attracts them along that direction.

**Applying surface potential in a finite-sized array**

Although there can be several methods to drive a surface potential into a photonic crystal, we herein adjust the surface terminations in a finite-sized array. In a fully periodic crystal, the fields or eigenvectors of a propagating mode are invariant under discrete translation, not depending on a specific position of the basis in a unit cell. On the contrary, a finite-sized array has terminations at the boundaries. A propagating wave profile depends on the terminations according to the variations of the basis's positions in the unit cells[84], as shown in Fig. 2a. Thus, the effect of surface potential can be quantitated with respect to adjusting the surface terminations.

Here, we remark on the followings: (i) Eqs. (1) to (4) do not consider a detailed geometry in each unit cell. Thus, observation of the relation between the surface termination and surface potential should be phenomenologically carried out using a real array structure. (ii) A photonic band structure for a finite-sized array displays projected bands, called folding of bands. Without the pseudomagnetic field, the zeroth Landau level adheres to the bulk bands so that surface states cannot be obtained. Thus, we should use the pseudomagnetic field, as shown in Fig. 2b. Our study shows the interplay between the surface potential and pseudomagnetic field; the pseudomagnetic field gives the surface localization of a photonic wave, and the surface potential tunes the degree of the localization.

To build the array in Fig. 2b and to induce both the surface potential and pseudomagnetic field, we use a DG photonic crystal that exhibits Weyl points[17]. We adjust the surface terminations by tuning the translation of the basis along the normal direction to the boundaries.

**Double gyroid photonic array exhibiting pseudomagnetic field**

To drive the surface potential by the surface termination, we use a photonic array that exhibits a pseudomagnetic field by $\mathbf{k}_w = \mathbf{k}_{w,0} + \mathbf{A}_w^n$ in eq. (4). We consider the DG, as shown in Fig. 3a. The yellow and blue single gyroids (SGs) are given by a set of $\mathbf{x} = [x_1, x_2, x_3]$ such that $f_{SG,Y}(\mathbf{x}) + pf_p(\mathbf{x}) > f_{D_2} > 0$ and $f_{SG,B}(-\mathbf{x}) > f_O > 0$, respectively, where $f_{D_2}$ and $f_O$ are the level-set values that determine the volume fraction of each SG. The yellow SG has selectively the perturbation term $pf_p(\mathbf{x})$ to break the inversion symmetry and to generate Weyl points in momentum space. Due to this perturbation, the yellow SG exhibits a defect-like narrow region on the arm passing the $\mathbf{a}_1\mathbf{a}_2$ surface, as shown in Fig. 3a. The higher perturbation strength $p$ induces the deeper defect-like shape; this corresponds to the yellow circle's size in Fig. 1a-b and Fig. 2. (See Methods for the detailed explanations about the DG crystal.) When this DG photonic crystal is periodic along all three lattice vector directions, it can exhibit four Weyl points, as shown in Fig. 3b-c. The Weyl points $H_0$ and $N_0$ (marked with blue and red solid points in Fig. 3c, respectively) have positive and negative topological charges, respectively.

Varying $p$ of the DG shifts the positions of the Weyl points, as shown in Fig. 3d. We denote the moved positions of the Weyl points $N_0$ and $H_0$ in momentum space by $\mathbf{k}_w^N$ and $\mathbf{k}_w^H$, respectively. Increasing $p$ makes the Weyl points shift away from Γ-point on the single plane ((001)-plane). For the narrow range of $p$ around the central value $p_0$, the traces of all Weyl points exhibit linear shapes so that we can write their positions as $\mathbf{k}_w^H = \mathbf{k}_{w,0}^H + \hat{\mathbf{k}}_s^H \delta k^H$ and $\mathbf{k}_w^N = \mathbf{k}_{w,0}^N + \hat{\mathbf{k}}_s^N \delta k^N$ where $\hat{\mathbf{k}}_s^H$ and $\hat{\mathbf{k}}_s^N$ are overall directions of the traces (see Fig. 3d). The deviations of shifted Weyl points, $\delta k^N$ and $\delta k^H$, also show linear relations with $p$, as shown in Fig. 3e.

To generate the pseudomagnetic field, we design a photonic array made of DGs whose $p$ linearly varies with respect to the position in the array[13,70,71,85,86]. First, we assume a DG-array that consists of DGs with $N$ unit cells along the $\mathbf{a}_\perp$-direction between two boundaries, as shown in Fig. 4a. The array is periodic along the $\mathbf{a}_=$- and $\mathbf{a}_\parallel$-directions. Next, we apply the non-uniform geometry on this array using the perturbation strength $p$ that linearly varies along the $\mathbf{a}_\perp$:

$$p = p_s\{x_\perp - d_0\} + p_0, \qquad (5)$$

like the inset linear plot in Fig. 4a. Here, $x_\perp$ is a coordinate along the $\mathbf{a}_\perp$, $d_0$ is a distance between the planes at $n = 0$, and $n = N/2$, and $p_0$ is a central value of $p(\mathbf{x})$. From the information in Fig. 4a-b, the pseudomagnetic field has the form $\mathbf{B} = \nabla_\mathbf{x} \times \mathbf{A} = p_s(B_= \hat{\mathbf{a}}_= + B_\parallel \hat{\mathbf{a}}_\parallel)$, which are parallel to the array boundaries (see Fig. 4c-d). As a result, the waves can be localized at the boundaries (see Methods for detailed explanations about the array and pseudomagnetic field calculations).

**Surface potential by surface termination and resulting photonic wave localization**

Plugging the design in the previous section to the array of $N = 48$ primitive cells allows us to generate Landau spectra. The detailed explanations of the Landau levels related to the pseudomagnetic field are in Section 1, Supplementary Information.

Here, we extract and compare the zeroth Landau levels by two arrays with $p_s = -2p_{s,0}$ and $p_s = +2p_{s,0}$, as illustrated in Fig. 5a. Although the two arrays have different internal geometries due to the different $p_s$ values, their overall translation status are identical, i.e., they generally use the formulae denoted in Fig. 3a. Thus, we consider that their surface terminations are identical. The photonic band structure in Fig. 5a shows that the resulting zeroth Landau levels accessible from the two arrays are not equal. Furthermore, the eigenstate intensity distributions for the two cases reveal mutual asymmetric (Fig. 5b-c). The distributions in Fig. 5d-e are biased toward the $n = 48$ from the symmetric curve (the gray dotted lines). From the fact that the results in Fig. 5d-e exhibit the same tendency as with Fig. 1e, we can conclude that there exists a surface potential in these arrays.

Now, we replace $\mathbf{x}$ as $\mathbf{x} - h\mathbf{a}_\perp$ to apply overall translation $h$ of the DG by $h|\mathbf{a}_\perp|$ along the $\mathbf{a}_\perp$-direction to verify the surface termination's effect, as illustrated in Fig. 6a. Note that the profile of $p$ does not move (see the upper right plot in Fig. 6a). For a specific point on the zeroth Landau levels (marked in Fig. 6b), we calculate the field intensity for several $h$ values, as shown in Fig. 6c. The discussions so far are about that the surface potential can adjust the degree of localization of surface states generated by the pseudomagnetic field. The results in Fig. 6c indicate that the localization degree can be adjusted by the surface termination. The zeroth Landau level in the photonic band structure also can be moved to another position by changing $h$. Due to the relatively small $h$, the overall configurations of the zeroth Landau level and the surrounding bulk bands remains almost intact, as shown in Fig. 6b.

**Evasion behavior of photonic wave**

We embody the evasion behavior of a photonic wave in a DG structure using oppositely graded $p$. Adjusting the propagation path of a photonic or phononic wave has been an interest in photonics or wave mechanics. These have been proved using materials with negative-refractive indices[16,87-89] or Dirac/Weyl crystals exhibiting one-way propagations of waves[3,12,13,44,90]. Such proofs are related to invisibility-cloaking. Here, we perform the behavior by pulling a photonic wave that was originally propagating on one boundary toward the opposite boundary.

DG arrays that consist of 8 cells with $p_s = -2p_{s,0}$, 0, $p_{s,0}$, $3p_{s,0}$, and $5p_{s,0}$ are prepared, based on the array in Fig. 4a, ($p_{s,0} > 0$). (The number of cells is counted based on the BCC primitive cell.) The reason for using a smaller number of cells than in the previous case is to broaden intervals between Landau levels. We classify these arrays into groups with $p_s = -2p_{s,0}$ and the others with $p_s \geq 0$. The directions of $\nabla_\mathbf{x} p$ for these two groups are opposite to each other (see the inset of Fig. 7a and b). The band structures for $p_s < 0$ and $p_s \geq 0$ (plotted in Fig. 7a and b, respectively) exhibit Landau levels along $\Gamma H'$-direction. From these plots, we consider $\omega a/2\pi c = 0.49$ as an optimized frequency for the evasion behavior because this frequency between the bulk bands and all zeroth Landau levels commonly meets this frequency only once. We then constitute a DG array, as shown in Fig. 7c, using the 8-cells arrays depicted in the inset of Fig. 7a-b. Each system consists of 8 and 32 cells along $\mathbf{a}_\perp$- and $\mathbf{a}_=$-directions, respectively, and it is assumed to be infinitely periodic along $\hat{\mathbf{a}}_\parallel$-direction. The blocks $m = [0, 8]$ and $m = [24, 32]$ use negative-valued $p_s$, and the central region between the two sections has positive-valued $p_s$.

The DG array reveals the evasion behavior of photonic waves. At $\omega a/2\pi c = 0.49$, a photonic wave is localized in the section of $p_s = -2p_{s,0}$ around the boundary at $n = 0$. We then input an incident photonic wave around the localization region, as marked by the star symbols in Fig. 7d-g. If the blocks $m = [8, 24]$ were the same as the other blocks, the localization would continue in the central region. We simulate here the situations that the central region's $p_s$ is respectively 0, $p_{s,0}$, $3p_{s,0}$, and $5p_{s,0}$. The results show the gradual attraction of the photonic waves onto the boundary at $n = 8$ with increasing $p_s$ (see Fig. 7d-g).

In addition, we observe the attraction of waves towards the $n = 0$ boundaries. In the blocks $m = [0, 8]$ and $m = [24, 32]$, the waves are localized on the boundary. On the contrary, the localized waves in the blocks $m = [8, 24]$ exhibit biased behaviors. Then, we can conclude that this arises from surface potential and surface termination. If we adjust the surface termination, we may observe the bias towards the opposite direction. All these show the interplay between surface potential and pseudomagnetic field.

**Discussion**

We have demonstrated the asymmetric localization of photonic waves via an interplay between surface potential and pseudomagnetic field using a DG photonic crystal. The pseudomagnetic field has induced the surface states of the photonic waves and the surface potential adjusted the degree of localization. The pseudomagnetic field was formed by the graded location of Weyl points, and the surface potential was applied using surface termination by tuning unit cells' basis translation. We have observed the shift of the surface states as a result of the surface potential.

In the case of graphene[65], boron nitride[66], and semimetals[60,64,67], surface potentials can be applied using materials' surface/edge passivation. Although this study used surface termination by crystal's basis translation, we believe that there could be several types of photonic passivation, for example, a thin material doping. Then, this study and the follow-up studies will open the various possibility of using topological photonic crystals. Meanwhile, there is no example of the detailed analysis of the surface states along the zeroth Landau level by the Hall effect and pseudomagnetic field in three-dimensional Weyl photonic crystals. Therefore, this study will fill this gap thereby this study will positively affect the other studies on the three-dimensional quantum (spin) Hall effect[91-96].

**References**

1       Qi, X.-L. & Zhang, S.-C. Topological insulators and superconductors. *Reviews of Modern Physics* **83**, 1057-1110 (2011).
2       Hasan, M. Z. & Kane, C. L. Colloquium: Topological insulators. *Reviews of Modern Physics* **82**, 3045-3067 (2010).
3       Slobozhanyuk, A. et al. Three-dimensional all-dielectric photonic topological insulator. *Nature Photonics* **11**, 130-136 (2017).
4       Lu, L. et al. Symmetry-protected topological photonic crystal in three dimensions. *Nature Physics* **12**, 337-340 (2016).
5       Jin, D. et al. Infrared Topological Plasmons in Graphene. *Physical Review Letters* **118**, 245301 (2017).
6       Liu, G.-G. et al. Observation of an unpaired photonic Dirac point. *Nature*

   *Communications* **11**, 1873 (2020).
7 Peng, B., Bouhon, A., Monserrat, B. & Slager, R.-J. Phonons as a platform for non-Abelian braiding and its manifestation in layered silicates. *Nature Communications* **13**, 423 (2022).
8 Jiang, B. et al. Experimental observation of non-Abelian topological acoustic semimetals and their phase transitions. *Nature Physics* (2021).
9 Peng, B., Bouhon, A., Slager, R.-J. & Monserrat, B. Multigap topology and non-Abelian braiding of phonons from first principles. *Physical Review B* **105**, 085115 (2022).
10 Lu, L. et al. Experimental observation of Weyl points. *Science* **349**, 622-624 (2015).
11 Yang, B. et al. Ideal Weyl points and helicoid surface states in artificial photonic crystal structures. *Science* **359**, 1013-1016 (2018).
12 Yang, Y. et al. Realization of a three-dimensional photonic topological insulator. *Nature* **565**, 622-626 (2019).
13 Jia, H. et al. Observation of chiral zero mode in inhomogeneous three-dimensional Weyl metamaterials. *Science* **363**, 148-151 (2019).
14 Sie, E. J. et al. An ultrafast symmetry switch in a Weyl semimetal. *Nature* **565**, 61-66 (2019).
15 Soluyanov, A. A. et al. Type-II Weyl semimetals. *Nature* **527**, 495 (2015).
16 He, H. et al. Topological negative refraction of surface acoustic waves in a Weyl phononic crystal. *Nature* **560**, 61-64 (2018).
17 Park, H. & Lee, S. Double Gyroids for Frequency-Isolated Weyl Points in the Visible Regime and Interference Lithographic Design. *ACS Photonics* **7**, 1577-1585 (2020).
18 Lu, L., Fu, L., Joannopoulos, J. D. & Soljačić, M. Weyl points and line nodes in gyroid photonic crystals. *Nature Photonics* **7**, 294 (2013).
19 Park, H. et al. Block copolymer gyroids for nanophotonics: significance of lattice transformations. *Nanophotonics* **11**, 2583-2615 (2022).
20 Ahn, J., Kim, D., Kim, Y. & Yang, B.-J. Band Topology and Linking Structure of Nodal Line Semimetals with Z2 Monopole Charges. *Physical Review Letters* **121**, 106403 (2018).
21 Wu, Q., Soluyanov, A. A. & Bzduŝek, T. Non-Abelian band topology in noninteracting metals. *Science* **365**, 1273-1277 (2019).
22 Xia, L. et al. Observation of Hourglass Nodal Lines in Photonics. *Physical Review Letters* **122**, 103903 (2019).
23 Tiwari, A. & Bzduŝek, T. Non-Abelian topology of nodal-line rings in PT-symmetric systems. *Physical Review B* **101**, 195130 (2020).
24 Kim, M., Jacob, Z. & Rho, J. Recent advances in 2D, 3D and higher-order topological photonics. *Light: Science & Applications* **9**, 130 (2020).
25 Park, H., Wong, S., Zhang, X. & Oh, S. S. Non-Abelian Charged Nodal Links in a Dielectric Photonic Crystal. *ACS Photonics* **8**, 2746-2754 (2021).
26 Park, H., Gao, W., Zhang, X. & Oh, S. S. Nodal lines in momentum space: topological invariants and recent realizations in photonic and other systems. *Nanophotonics* **11**, 2779-2801 (2022).
27 Park, H. & Oh, S. S. Sign freedom of non-abelian topological charges in phononic and photonic topological semimetals. *New Journal of Physics* **24**, 053042 (2022).
28 Park, H., Wong, S., Bouhon, A., Slager, R.-J. & Oh, S. S. Topological phase transitions of non-Abelian charged nodal lines in spring-mass systems. *Physical Review B* **105**, 214108 (2022).
29 Xie, Y., Cai, J., Kim, J., Chang, P.-Y. & Chen, Y. Hopf-chain networks evolved from

| | triple points. *Physical Review B* **99**, 165147 (2019). |
|---|---|
| 30 | Nielsen, H. B. & Ninomiya, M. The Adler-Bell-Jackiw anomaly and Weyl fermions in a crystal. *Physics Letters B* **130**, 389-396 (1983). |
| 31 | Son, D. T. & Spivak, B. Z. Chiral anomaly and classical negative magnetoresistance of Weyl metals. *Physical Review B* **88**, 104412 (2013). |
| 32 | Huang, X. et al. Observation of the Chiral-Anomaly-Induced Negative Magnetoresistance in 3D Weyl Semimetal TaAs. *Physical Review X* **5**, 031023 (2015). |
| 33 | Yang, K.-Y., Lu, Y.-M. & Ran, Y. Quantum Hall effects in a Weyl semimetal: Possible application in pyrochlore iridates. *Physical Review B* **84**, 075129 (2011). |
| 34 | Wan, X., Turner, A. M., Vishwanath, A. & Savrasov, S. Y. Topological semimetal and Fermi-arc surface states in the electronic structure of pyrochlore iridates. *Physical Review B* **83**, 205101 (2011). |
| 35 | Burkov, A. A. & Balents, L. Weyl Semimetal in a Topological Insulator Multilayer. *Physical Review Letters* **107**, 127205 (2011). |
| 36 | Potter, A. C., Kimchi, I. & Vishwanath, A. Quantum oscillations from surface Fermi arcs in Weyl and Dirac semimetals. *Nature Communications* **5**, 5161 (2014). |
| 37 | Peri, V., Serra-Garcia, M., Ilan, R. & Huber, S. D. Axial-field-induced chiral channels in an acoustic Weyl system. *Nature Physics* **15**, 357-361 (2019). |
| 38 | Xiao, M., Chen, W.-J., He, W.-Y. & Chan, C. T. Synthetic gauge flux and Weyl points in acoustic systems. *Nature Physics* **11**, 920 (2015). |
| 39 | Li, F., Huang, X., Lu, J., Ma, J. & Liu, Z. Weyl points and Fermi arcs in a chiral phononic crystal. *Nature Physics* **14**, 30 (2017). |
| 40 | Lee, C. H. et al. Imaging nodal knots in momentum space through topolectrical circuits. *Nature Communications* **11**, 4385 (2020). |
| 41 | Wang, L., Jian, S.-K. & Yao, H. Topological photonic crystal with equifrequency Weyl points. *Physical Review A* **93**, 061801 (2016). |
| 42 | Goi, E., Yue, Z., Cumming, B. P. & Gu, M. Observation of Type I Photonic Weyl Points in Optical Frequencies. *Laser & Photonics Reviews* **12**, 1700271 (2018). |
| 43 | Yang, Z. et al. Weyl points in a magnetic tetrahedral photonic crystal. *Opt. Express* **25**, 15772-15777 (2017). |
| 44 | Lu, L., Gao, H. & Wang, Z. Topological one-way fiber of second Chern number. *Nature Communications* **9**, 5384 (2018). |
| 45 | Fruchart, M. et al. Soft self-assembly of Weyl materials for light and sound. *Proceedings of the National Academy of Sciences of the United States of America* **115**, E3655-E3664 (2018). |
| 46 | Yang, Y. et al. Ideal Unconventional Weyl Point in a Chiral Photonic Metamaterial. *Physical Review Letters* **125**, 143001 (2020). |
| 47 | Noh, J. et al. Experimental observation of optical Weyl points and Fermi arc-like surface states. *Nature Physics* **13**, 611-617 (2017). |
| 48 | He, H. et al. Observation of quadratic Weyl points and double-helicoid arcs. *Nature Communications* **11**, 1820 (2020). |
| 49 | Lu, L., Joannopoulos, J. D. & Soljačić, M. Topological photonics. *Nature Photonics* **8**, 821 (2014). |
| 50 | Gong, Y., Wong, S., Bennett, A. J., Huffaker, D. L. & Oh, S. S. Topological Insulator Laser Using Valley-Hall Photonic Crystals. *ACS Photonics* **7**, 2089-2097 (2020). |
| 51 | Wong, S., Saba, M., Hess, O. & Oh, S. S. Gapless unidirectional photonic transport using all-dielectric kagome lattices. *Physical Review Research* **2**, 012011 (2020). |
| 52 | Deng, W. et al. Nodal rings and drumhead surface states in phononic crystals. *Nature Communications* **10**, 1769 (2019). |

53  Gao, W. et al. Experimental observation of photonic nodal line degeneracies in metacrystals. *Nature Communications* **9**, 950 (2018).
54  Chang, G. et al. Topological Hopf and Chain Link Semimetal States and Their Application to Co2MnGa. *Physical Review Letters* **119**, 156401 (2017).
55  Belopolski, I. et al. Discovery of topological Weyl fermion lines and drumhead surface states in a room temperature magnet. *Science* **365**, 1278-1281 (2019).
56  Guo, Q. et al. Observation of Three-Dimensional Photonic Dirac Points and Spin-Polarized Surface Arcs. *Physical Review Letters* **122**, 203903 (2019).
57  Yang, Y. et al. Observation of a topological nodal surface and its surface-state arcs in an artificial acoustic crystal. *Nature Communications* **10**, 5185 (2019).
58  Xu, S.-Y. et al. Discovery of a Weyl fermion state with Fermi arcs in niobium arsenide. *Nature Physics* **11**, 748-754 (2015).
59  Xu, S.-Y. et al. Discovery of a Weyl fermion semimetal and topological Fermi arcs. *Science* **349**, 613-617 (2015).
60  Potter, A. C., Kimchi, I. & Vishwanath, A. Quantum oscillations from surface Fermi arcs in Weyl and Dirac semimetals. *Nature Communications* **5**, 5161 (2014).
61  Bhowmick, S. & Shenoy, V. B. Weber-Fechner type nonlinear behavior in zigzag edge graphene nanoribbons. *Physical Review B* **82**, 155448 (2010).
62  Tchoumakov, S., Civelli, M. & Goerbig, M. O. Magnetic description of the Fermi arc in type-I and type-II Weyl semimetals. *Physical Review B* **95**, 125306 (2017).
63  Fefferman, C. L., Lee-Thorp, J. P. & Weinstein, M. I. Edge States in Honeycomb Structures. *Annals of PDE* **2**, 12 (2016).
64  Dongre, N. K. & Roychowdhury, K. Effects of surface potentials on Goos-Haenchen and Imbert-Fedorov shifts in Weyl semimetals. *arXiv preprint arXiv:2106.04573* (2021).
65  Shtanko, O. & Levitov, L. Robustness and universality of surface states in Dirac materials. *Proceedings of the National Academy of Sciences* **115**, 5908-5913 (2018).
66  Dou, Z. et al. Imaging Bulk and Edge Transport near the Dirac Point in Graphene Moiré Superlattices. *Nano Letters* **18**, 2530-2537 (2018).
67  Sun, Y., Wu, S.-C. & Yan, B. Topological surface states and Fermi arcs of the noncentrosymmetric Weyl semimetals TaAs, TaP, NbAs, and NbP. *Physical Review B* **92**, 115428 (2015).
68  Ilan, R., Grushin, A. G. & Pikulin, D. I. Pseudo-electromagnetic fields in 3D topological semimetals. *Nature Reviews Physics* **2**, 29-41 (2020).
69  Guinea, F., Katsnelson, M. I. & Geim, A. K. Energy gaps and a zero-field quantum Hall effect in graphene by strain engineering. *Nature Physics* **6**, 30 (2010).
70  Abbaszadeh, H., Souslov, A., Paulose, J., Schomerus, H. & Vitelli, V. Sonic Landau Levels and Synthetic Gauge Fields in Mechanical Metamaterials. *Physical Review Letters* **119**, 195502 (2017).
71  Brendel, C., Peano, V., Painter, O. J. & Marquardt, F. Pseudomagnetic fields for sound at the nanoscale. *Proceedings of the National Academy of Sciences*, 201615503 (2017).
72  Levy, N. et al. Strain-Induced Pseudo–Magnetic Fields Greater Than 300 Tesla in Graphene Nanobubbles. *Science* **329**, 544-547 (2010).
73  Klitzing, K. v., Dorda, G. & Pepper, M. New Method for High-Accuracy Determination of the Fine-Structure Constant Based on Quantized Hall Resistance. *Physical Review Letters* **45**, 494-497 (1980).
74  Hall, E. H. On a New Action of the Magnet on Electric Currents. *American Journal of Mathematics* **2**, 287-292 (1879).
75  Shi, H. et al. Large-area, periodic, and tunable intrinsic pseudo-magnetic fields in low-

|     | angle twisted bilayer graphene. *Nature Communications* **11**, 371 (2020). |
| --- | --- |
| 76  | Grushin, A. G., Venderbos, J. W. F., Vishwanath, A. & Ilan, R. Inhomogeneous Weyl and Dirac Semimetals: Transport in Axial Magnetic Fields and Fermi Arc Surface States from Pseudo-Landau Levels. *Physical Review X* **6**, 041046 (2016). |
| 77  | Pikulin, D. I., Chen, A. & Franz, M. Chiral Anomaly from Strain-Induced Gauge Fields in Dirac and Weyl Semimetals. *Physical Review X* **6**, 041021 (2016). |
| 78  | Liu, C.-X., Ye, P. & Qi, X.-L. Chiral gauge field and axial anomaly in a Weyl semimetal. *Physical Review B* **87**, 235306 (2013). |
| 79  | Heidari, S. & Asgari, R. Chiral Hall effect in strained Weyl semimetals. *Physical Review B* **101**, 165309 (2020). |
| 80  | Rechtsman, M. C. et al. Strain-induced pseudomagnetic field and photonic Landau levels in dielectric structures. *Nature Photonics* **7**, 153 (2012). |
| 81  | Wang, W. et al. Moiré Fringe Induced Gauge Field in Photonics. *Physical Review Letters* **125**, 203901 (2020). |
| 82  | Mann, C.-R., Horsley, S. A. R. & Mariani, E. Tunable pseudo-magnetic fields for polaritons in strained metasurfaces. *Nature Photonics* **14**, 669-674 (2020). |
| 83  | Wen, X. et al. Acoustic Landau quantization and quantum-Hall-like edge states. *Nature Physics* **15**, 352-356 (2019). |
| 84  | Joannopoulos, J. D., Johnson, S. G., Winn, J. N. & Meade, R. D. *Photonic Crystals: Molding the Flow of Light*. Second Edition edn, (Princeton University Press, 2008). |
| 85  | Lu, C., Wang, C., Xiao, M., Zhang, Z. Q. & Chan, C. T. Topological Rainbow Concentrator Based on Synthetic Dimension. *Physical Review Letters* **126**, 113902 (2021). |
| 86  | Chaplain, G. J., Pajer, D., De Ponti, J. M. & Craster, R. V. Delineating rainbow reflection and trapping with applications for energy harvesting. *New Journal of Physics* **22**, 063024 (2020). |
| 87  | Pendry, J. B. Negative Refraction Makes a Perfect Lens. *Physical Review Letters* **85**, 3966-3969 (2000). |
| 88  | Smith, D. R., Pendry, J. B. & Wiltshire, M. C. K. Metamaterials and Negative Refractive Index. *Science* **305**, 788-792 (2004). |
| 89  | Kaina, N., Lemoult, F., Fink, M. & Lerosey, G. Negative refractive index and acoustic superlens from multiple scattering in single negative metamaterials. *Nature* **525**, 77-81 (2015). |
| 90  | Wang, P., Lu, L. & Bertoldi, K. Topological Phononic Crystals with One-Way Elastic Edge Waves. *Physical Review Letters* **115**, 104302 (2015). |
| 91  | Tang, F. et al. Three-dimensional quantum Hall effect and metal–insulator transition in ZrTe5. *Nature* **569**, 537-541 (2019). |
| 92  | Lu, H.-Z. 3D quantum Hall effect. *National Science Review* **6**, 208-210 (2018). |
| 93  | Zhang, W. et al. Observation of a thermoelectric Hall plateau in the extreme quantum limit. *Nature Communications* **11**, 1046 (2020). |
| 94  | Zheng, C., Yang, K. & Wan, X. Thouless conductances of a three-dimensional quantum Hall system. *Physical Review B* **102**, 064208 (2020). |
| 95  | Wang, C., Gioia, L. & Burkov, A. A. Fractional Quantum Hall Effect in Weyl Semimetals. *Physical Review Letters* **124**, 096603 (2020). |
| 96  | Qin, F. et al. Theory for the Charge-Density-Wave Mechanism of 3D Quantum Hall Effect. *Physical Review Letters* **125**, 206601 (2020). |
| 97  | Ashby, P. E. C. & Carbotte, J. P. Magneto-optical conductivity of Weyl semimetals. *Physical Review B* **87**, 245131 (2013). |

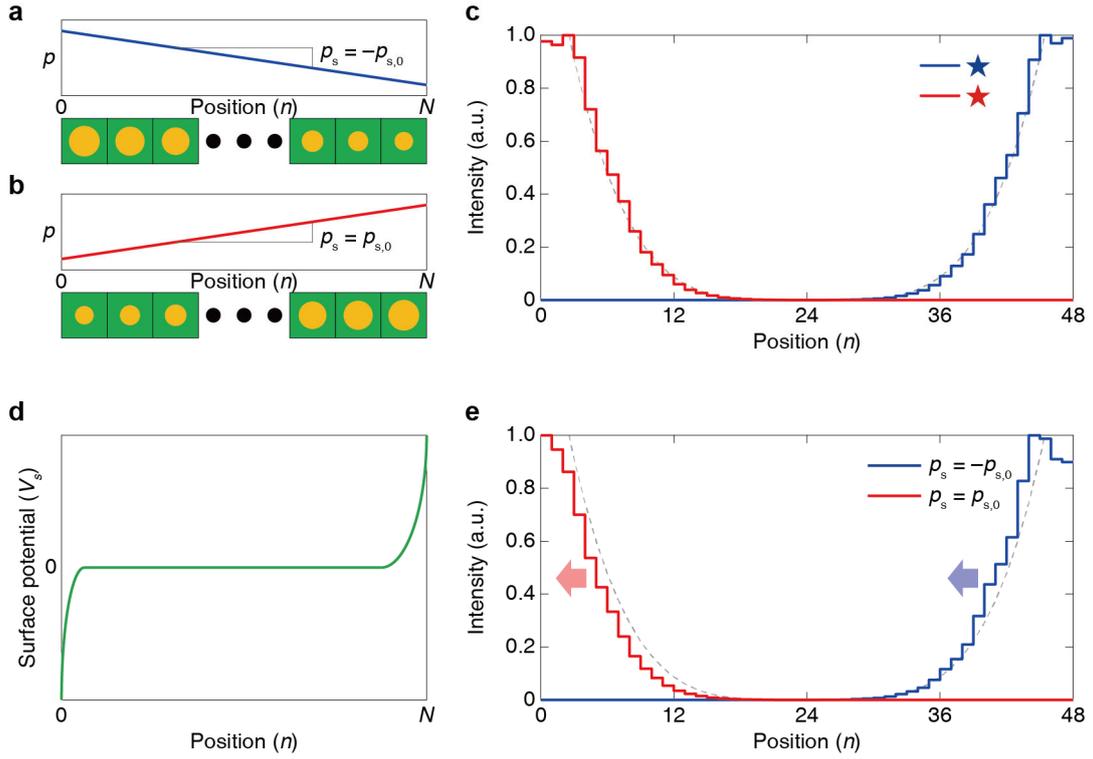

**Fig. 1. Effects of surface potential on the wave localization by pseudomagnetic fields. a**, **b**, Schematic illustrations of two systems with opposite $p_s$. The systems are finite along $x_3$-direction (the horizontal direction) and periodic along $x_1$- and $x_2$-direction. In each panel, the varying $p$ is schematically represented as the size of yellow circles. Both systems are described by equation (1). **c**, Comparisons of their eigenstates' distributions along $x_3$-direction, exhibiting the mutual symmetric dispersions. **d**, Schematic plot of $V_s$, the scalar coefficient of the surface potential. **e**, Comparisons of the eigenstates' distributions of **a** and **b** when $V_s$ is applied. In **c** and **e**, gray dotted lines are the symmetric curves with respect to the center for comparisons of the red and blue plots.

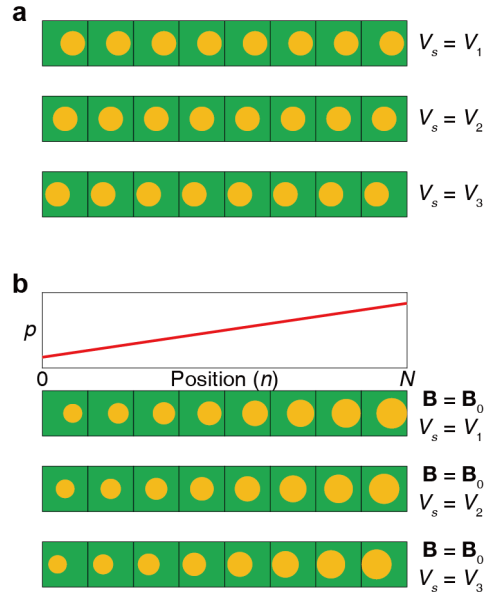

**Fig. 2. Schematics on the realization of surface potential. a**, Inducing different surface potentials by different surface terminations. The surface terminations are adjusted by the basis's positions in a unit cell. **b**, Adding pseudomagnetic field on the systems in **a**. Generating surface localization of a photonic wave is performed by the pseudomagnetic field, and the degree of localization depends on the surface potential.

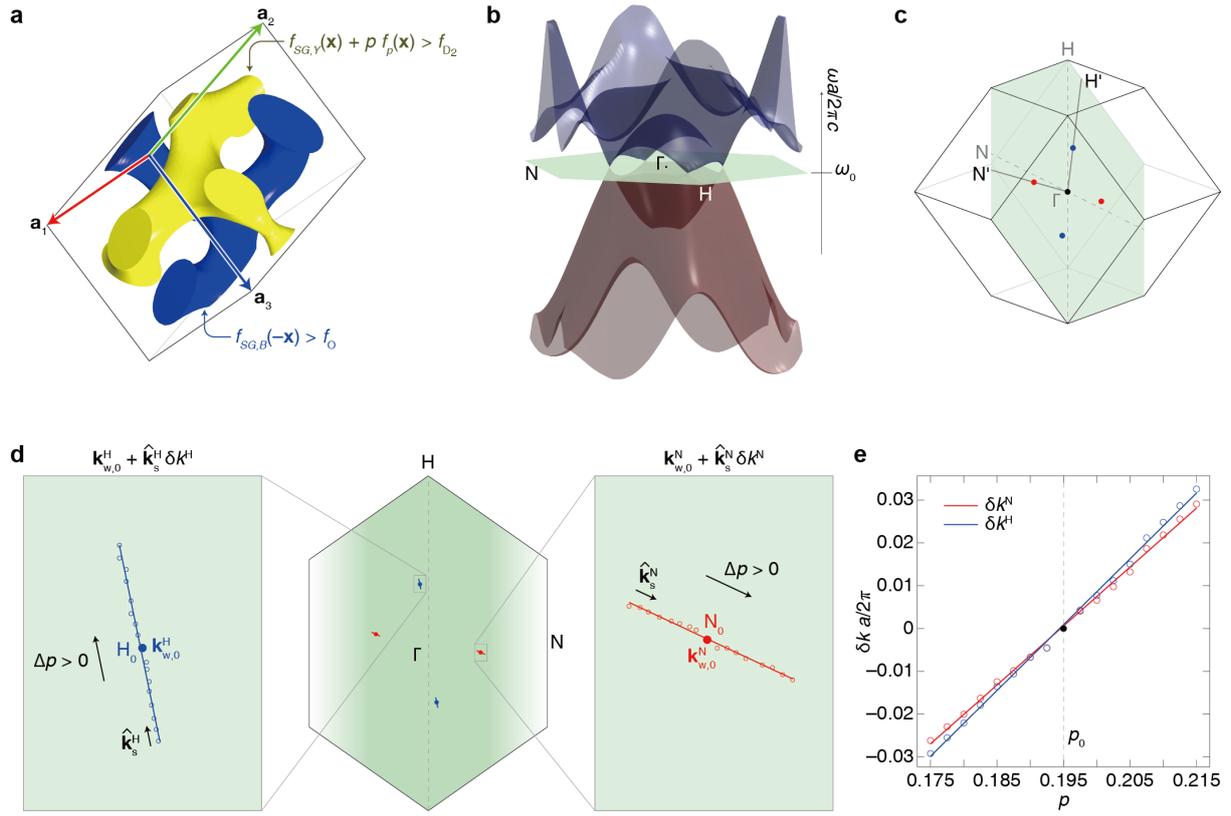

**Fig. 3. Design of 3D photonic crystal for pseudomagnetic field. a**, A DG photonic crystal. The SGs are defined by **x** that satisfies the inequalities denoted in the figure. The yellow SG's inequality has an additional term $pf_p(\mathbf{x})$ to impose a defect-like shape on this structure (see the bottom-right arm). **b**, Resulting photonic band structure exhibiting four Weyl points, the band degeneracies. **c**, Weyl points marked in the three-dimensional first Brillouin zone. They are on a single plane. **d**, Movements of Weyl points on the plane with the perturbation strength $p$. Weyl points in **c** are also marked here with red and blue solid points and denoted by $N_0$ and $H_0$, respectively. **e**, Deviations of the Weyl points' movements with $p$ from the positions of $N_0$ and $H_0$, shown in the right and left enlargements in **d**, respectively.

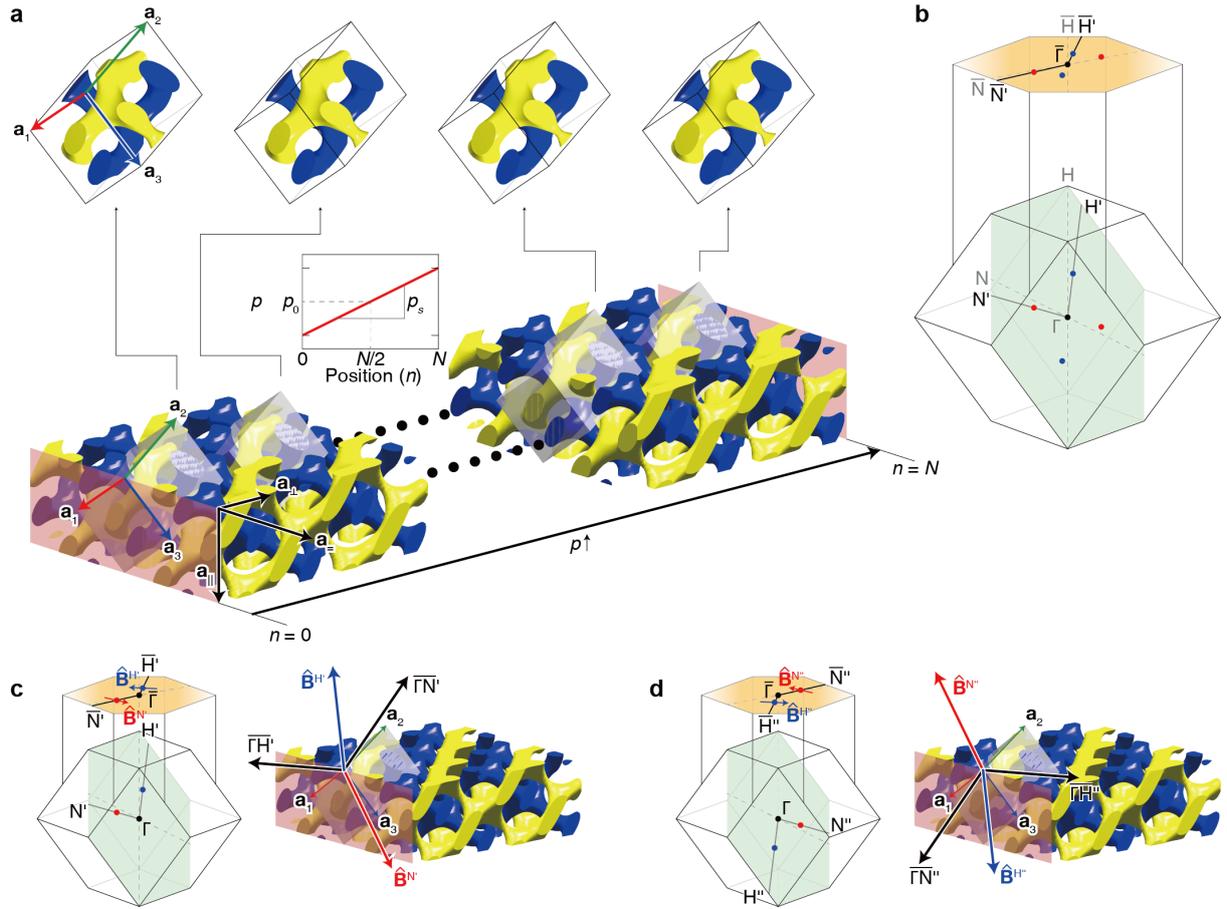

**Fig. 4. Design of a photonic array and resulting pseudomagnetic fields. a**, Schematic geometry which consists of $N$ cells with spatially linearly varying $p$ along $\mathbf{a}_\perp$-direction. This gradient is imposed only on the yellow structure. **b**, Bulk and surface Brillouin zones. **c-d**, Direction vectors of the pseudomagnetic field around each Weyl point by the above design, denoted by $\hat{\mathbf{B}}^{N'}$, $\hat{\mathbf{B}}^{H'}$, $\hat{\mathbf{B}}^{N''}$, and $\hat{\mathbf{B}}^{H''}$, respectively. These are calculated from the traces around $N_0$, $H_0$, $-N_0$, and $-H_0$ marked in Fig. 3d, respectively.

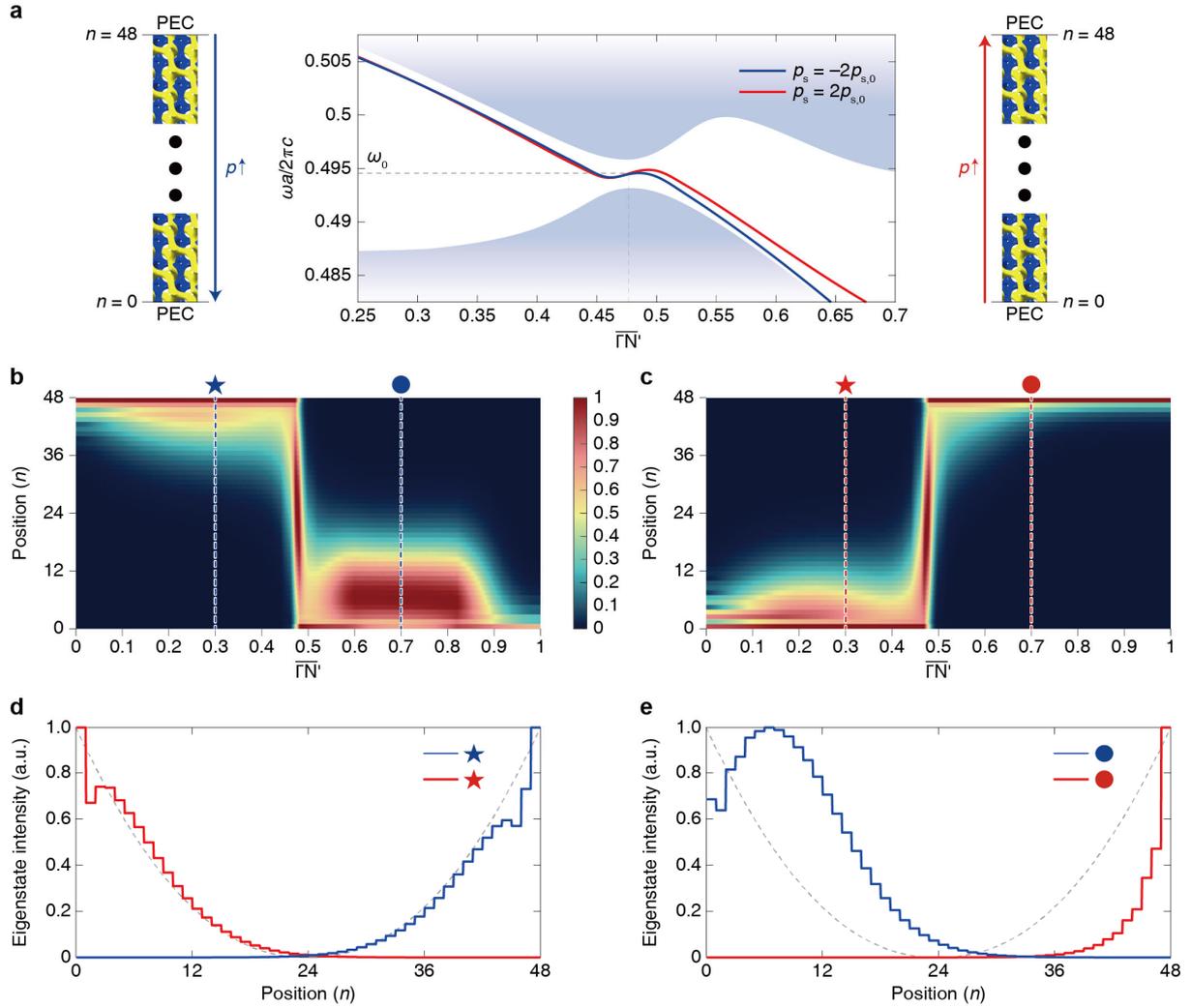

**Fig. 5. Photonic wave localization and evidence of surface potential. a**, Comparison of Landau levels by the opposite perturbation fields. The left and right cases correspond to $p_s = -2p_{s,0}$ and $p_s = 2p_{s,0}$, respectively. **b**, **c**, Normalized eigenstates respectively along the blue- and red-colored Landau levels shown in **a** to see surface states localized on the boundaries. **d**, **e**, Comparisons of the wave localization along the vertical lines in **b** and **c**. Data with the same symbols are overlapped in the same plot. Gray dotted lines are the symmetric curves with respect to $n = 24$ for comparisons of the red and blue plots.

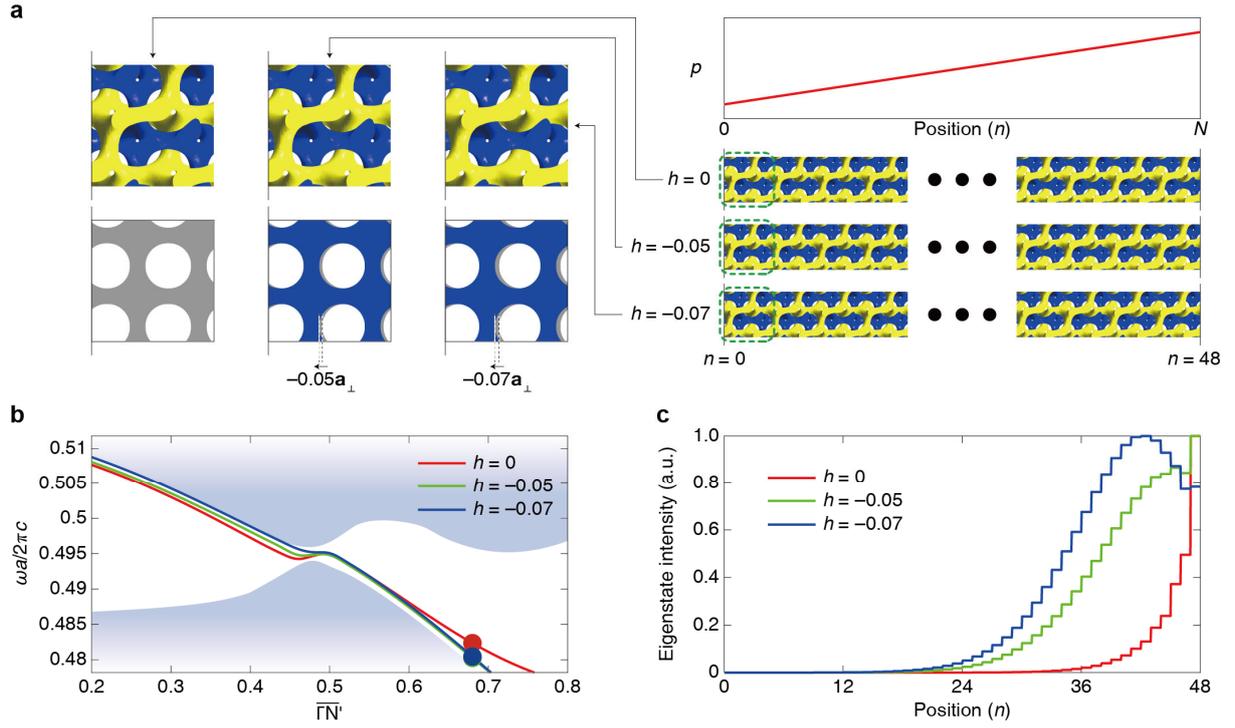

**Fig. 6. Adjustment of surface potential by surface termination. a**, Schematic illustration of three surface terminations. Observation of the surface potentials is carried out by adjusting the basis's positions in each unit cell. The upper left three structures are enlargements of the bottom right arrays. The lower left three figures are respectively the schematics of the upper left structures, and they clearly show the translation of the structures along $\mathbf{a}_\perp$-direction. The right upper plot indicates that the perturbation $p$ versus the position $n$ is the same for all three cases. **b-c**, Zeroth Landau levels (**b**) and eigenstate intensities (**c**) at the point marked in **b** for three different surface terminations tuned by translation $h$ along the $\mathbf{a}_\perp$-direction when $N = 48$ and $p_s = 2p_{s,0}$ are used.

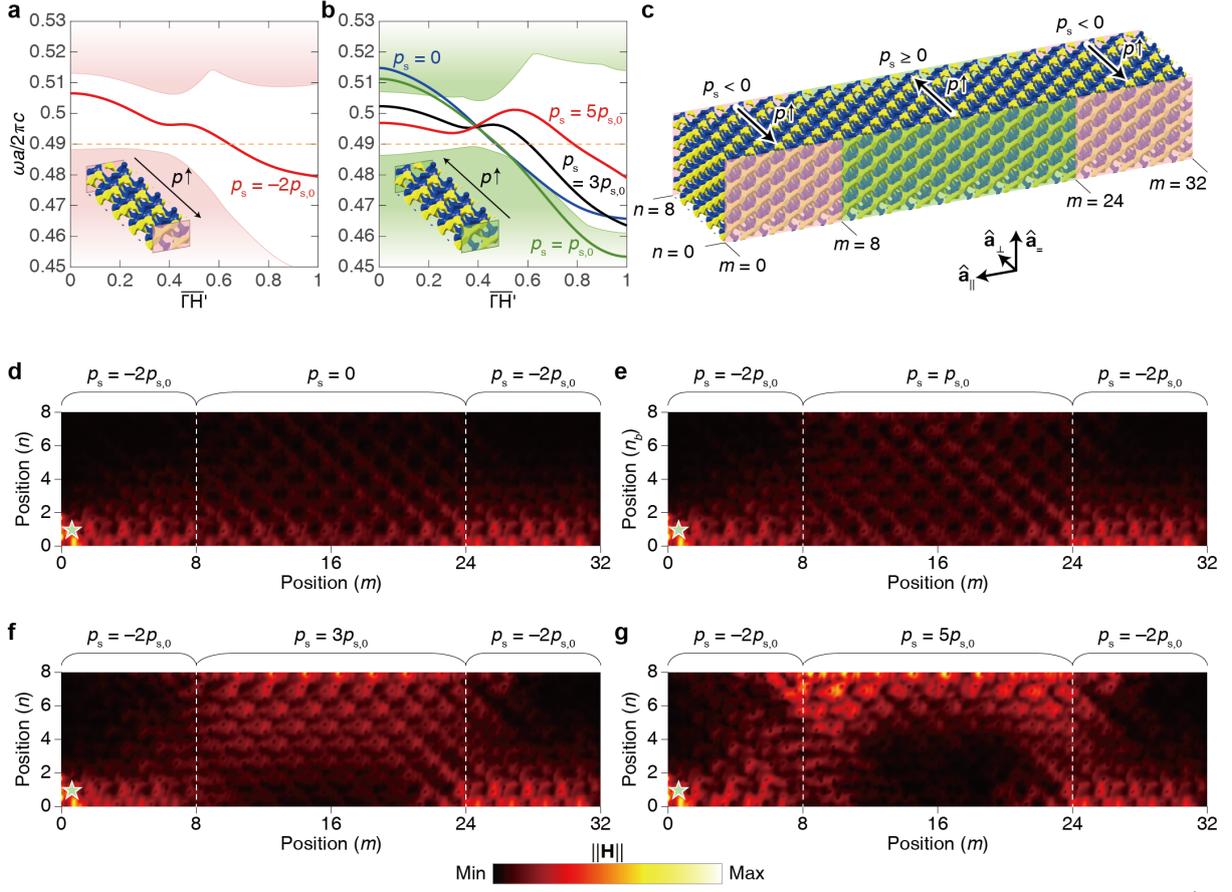

**Fig. 7. Evasion behavior of photonic waves using heterogeneous blocks. a**, **b**, Zeroth-Landau levels by the 8-block system with several perturbation fields, where $p_{s,0} = 7.0711 \times 10^{-3} a^{-1}$. The perturbation fields used in **a** and **b** are opposite. **c**, Heterogeneous DG system. The PEC is applied to the pink and green boundaries. In pink and green colored bounded blocks, the perturbation fields' directions are opposite as marked. The system is periodic only along $\hat{a}_\parallel$-direction. **d-g**, Evasion behavior of photonic waves with several $p_s > 0$ values of the central blocks with fixing $p_s < 0$ values of the blocks around both ends. Incident points are marked as the star symbols.

**Methods**

**Eigenstates localization by Weyl Hamiltonian.** We apply eq. (1) or (4) to the finite system illustrated in Fig. 1a-b to observe the wave localization like Fig. 1c or e. Instead of deriving the Landau level analytically using ladder operators[33,97], we use a traveling wave solution $\psi(\mathbf{x}, t) = u e^{i(\mathbf{k} \cdot \mathbf{x} - \omega t)}$ to consider the finite array. The traveling wave solution can be rewritten as the product of a state $u_n$ that depends on only $n$ and an exponent $e^{i(k_1 x_1 + k_2 x_2 - \omega t)}$ that depends on only other variables: $\psi(\mathbf{x}, t) = u_n e^{i(k_1 x_1 + k_2 x_2 - \omega t)}$. The surface localization is obtained by substituting this into equation (1) or (4). To consider the surface potential in eq. (3), we used the plot in Fig. 8 as $V_s$. Detailed derivations, explanations, and additional results related to Fig. 1 are given in Section 2, Supplementary Information.

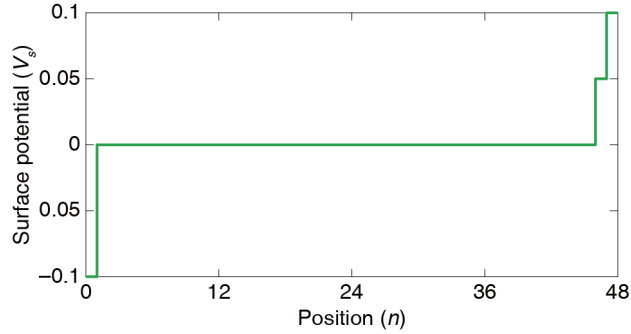

**Fig. 8. Plot of the surface potential's scalar coefficient $V_s$.** This is the discretized function from Fig. 1d. Plots in Fig. 1c and e were obtained using this $V_s$.

**Preparations of DG photonic crystal.** In the following, we give detailed explanations of the DG photonic crystal and array. In this study, we consider the DG, reported in ref. [17]. The body-centered cubic (BCC) primitive cell of this structure is defined by the lattice vectors $\mathbf{a}_i = a/2\,[1,1,1] - a\hat{\mathbf{x}}_i$, where $[\hat{\mathbf{x}}_1 \hat{\mathbf{x}}_2 \hat{\mathbf{x}}_3] = I$. The mathematical formulae of $f_{SG,Y}(\mathbf{x})$, $f_p(\mathbf{x})$, and $f_{SG,B}(\mathbf{x})$ denoted in Fig. 3a are $f_{SG,Y}(\mathbf{x}) = \sin X_1 \cos X_2 + \sin X_2 \cos X_3 + \sin X_3 \cos X_1$, $f_p(\mathbf{x}) = \sin(X_1 + X_2)$, and $f_{SG,B}(\mathbf{x}) = \sin \tilde{X}_1 \cos \tilde{X}_2 + \sin \tilde{X}_2 \cos \tilde{X}_3 + \sin \tilde{X}_3 \cos \tilde{X}_1$, respectively. Here, the local coordinates are given by $\mathbf{X} = [X_1, X_2, X_3] = (2\pi/a)(\mathbf{x} - \mathbf{a}_s s)$ and $\tilde{\mathbf{X}} = [\tilde{X}_1, \tilde{X}_2, \tilde{X}_3] = (2\pi/a)\mathbf{x}$, where $a$ is a lattice constant, $s = 0.0578$ is the shift coefficient that describes the translation of the yellow SG, and $\mathbf{a}_s = \mathbf{a}_1 + \mathbf{a}_2 + 2\mathbf{a}_3$ is the translation direction. For the inequalities marked in Fig. 3a, the level-set values $f_{D_2} = 1.15$ and $f_O = 1.1$ are used that determine the volume fraction of each SG. The refractive indices of the two SGs are commonly 4.0, and the outside region of them is filled with air whose refractive index is 1.0.

The perturbation strength $p$ plays an important role in our work. If $p$ becomes zero, the space group of this DG is $Ia\bar{3}d$ (no. 230), and this does not exhibit Weyl points[18]. A DG with appropriate nonzero-valued $p$ exhibits four Weyl points in the momentum space between the fourth and fifth bands[17]. Due to the inversion symmetry of the momentum space, the positions of these four Weyl points are inversion symmetric to $\Gamma$-point. When $p = p_0 = 0.195$ is used, we can get the four Weyl points, as shown in Fig. 3b-c. They are on a single (001)-plane passing $\Gamma N = (-\mathbf{b}_1 + \mathbf{b}_2)/2$ and $\Gamma H = (\mathbf{b}_1 + \mathbf{b}_2 - \mathbf{b}_3)/2$ where the reciprocal primitive vectors $\mathbf{b}_i$ are defined by $[\mathbf{b}_1 \mathbf{b}_2 \mathbf{b}_3]^T = 2\pi[\mathbf{a}_1 \mathbf{a}_2 \mathbf{a}_3]^{-1}$. These Weyl points are also denoted as $N_0$ and $H_0$ in Fig. 3d. The Chern numbers of the Weyl points $N_0$ and $H_0$ are $-1$ and $+1$, respectively (see Section 3, Supplementary Information).

**DG photonic array for pseudomagnetic field.** First, we assume 48 DGs along the $\mathbf{a}_\perp$-direction between two parallel perfect electric conductor (PEC) boundaries, where $\mathbf{a}_\perp = \mathbf{a}_2 + 0.5(\mathbf{a}_1 + \mathbf{a}_3)$ is normal to the boundaries (see Fig. 4a). The array is periodic along $\mathbf{a}_=$- and $\mathbf{a}_\parallel$-directions, where $\mathbf{a}_= = -\mathbf{a}_1 + \mathbf{a}_3$ and $\mathbf{a}_\parallel = \mathbf{a}_1 + \mathbf{a}_3$. The boundaries are parallel to $\mathbf{a}_1$ and $\mathbf{a}_3$, while the (001)-plane shown in Fig. 4b is parallel to $\mathbf{a}_1$ and $\mathbf{a}_2$ (see Fig. S4 in Section 4, Supplementary Information). Thus, the boundaries and (001)-plane are neither parallel nor perpendicular. The Weyl points $N_0$ and $H_0$ on the (001)-plane is partially conserved by projecting it onto the boundary, as shown in Fig. 4b. Then, the perturbation strength is given by the position-dependent form, i.e., $p = p(\mathbf{x})$. It linearly varies with the distance from the boundary at $n = 0$ along the $\mathbf{a}_\perp$, i.e., $\nabla_\mathbf{x} p \propto \mathbf{a}_\perp$, and it equals $p_0$ at the midplane between the boundaries, as schematically illustrated in the inset of Fig. 4a. The DGs in the primitive cells in Fig. 4a exhibit different shapes. Especially, the yellow parts show a stronger defect-like shape with larger $p$.

To calculate Fig. 5 and Fig. 6, we use $p_s = \pm p_{s,0}$ where $p_{s,0} = 2.9463 \times 10^{-4} a^{-1}$. The details not mentioned here are the same as the explanations in Section 1, Supplementary Information.

**Pseudomagnetic field by DG photonic array.** Let us assume that the effective Hamiltonian around a Weyl point is expressed as $H_{eff} = \sum_{i,j}^{3} v_{ij}(k_i - k_w)\sigma_j$ where $\mathbf{k}_w$ is the Weyl points locations ($\mathbf{k}_w^H$ or $\mathbf{k}_w^N$), $v_{ij}$ is the velocity tensor, and $\boldsymbol{\sigma}_j$ are the Pauli matrices[13,15]. $\mathbf{k}_w$ can be decomposed into $\mathbf{k}_w = \mathbf{k}_{w,0} + \hat{\mathbf{k}}_s \delta k = \mathbf{k}_{w,0} + \mathbf{A}$ where the superscripts N or H of all terms are omitted. Only the last term $\mathbf{A}$ relies on the real space coordinate-dependent $p$, i.e., $\mathbf{A} = \mathbf{A}(p(\mathbf{x}))$. From the information in Fig. 4a-b, the resulting pseudomagnetic field is, therefore, written as

$$\mathbf{B} = \nabla_\mathbf{x} \times \mathbf{A} = p_s(B_= \hat{\mathbf{a}}_= + B_\parallel \hat{\mathbf{a}}_\parallel) \tag{6}$$

where $\hat{\mathbf{a}}_=$ and $\hat{\mathbf{a}}_\parallel$ are the unit vectors of $\mathbf{a}_=$ and $\mathbf{a}_\parallel$, respectively, placed on the boundary (see Fig. 4a). The components $B_=$ and $B_\parallel$ are determined by the trajectories in Fig. 3d-e. Meanwhile, the proportional constant $p_s$ is the gradient of geometrical non-uniformity and is defined by the ratio of the $p$ change to the distance between these two boundaries (see the inset in Fig. 4a). (Detailed derivations and explanations of this result are given in Section 4,

Supplementary Information.) Because the length between the two boundaries can be written with the lattice constant $a$, the proportional constant $p_s$ can be expressed in terms of $a^{-1}$.

The pseudomagnetic field $\mathbf{B}$ in equation (6) has the linear combination form of $\hat{\mathbf{a}}_=$ and $\hat{\mathbf{a}}_\parallel$, parallel to the boundaries, so the field is parallel to the boundaries, as marked in Fig. 4c-d. In other words, the common perpendicular direction of $\hat{\mathbf{a}}_=$ and $\hat{\mathbf{a}}_\parallel$ coincides with the surface normal to the boundaries. As a result, we can quantitate the overall Hall effect driven by the pseudomagnetic field $\mathbf{B}$ and the resultant wave localization around boundaries.

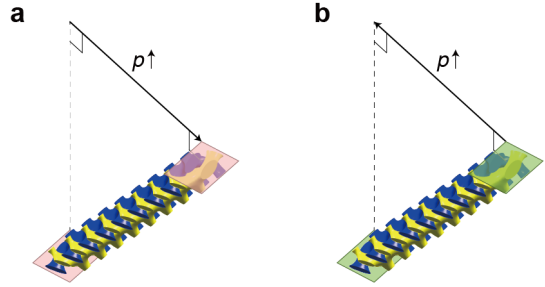

**Fig. 9. DG primitive cells array that was input to get Fig. 7a and b in the main text.** The array consists of 8 primitive cells with spatially linearly decreasing or increasing $p$ along $\hat{\mathbf{a}}_\perp$-direction.

**Evasion of the photonic wave.** Like the other photonic band structure calculations in this study, Fig. 7a-b is calculated using the array that consists of DG primitive cells, as illustrated in Fig. 9. The perturbation strength slope is counted by $p_{s,0} = 7.0711 \times 10^{-3} a^{-1}$, e.g., $p_s = 3p_{s,0}$. Fig. 7d-g are calculated using the array that consists of big cells whose lattice vectors are $2\mathbf{a}_\perp$, $\mathbf{a}_=$, and $\mathbf{a}_\parallel$, where $\mathbf{a}_\perp = (a/2)[1,0,1]$, $\mathbf{a}_= = -\mathbf{a}_1 + \mathbf{a}_3 = a[1,0,-1]$, and $\mathbf{a}_\parallel = \mathbf{a}_1 + \mathbf{a}_3 = a[0,1,0]$, respectively. We regard that the one big cell coincides with two DG primitive cells because of the projection of $\mathbf{a}_2$ and $\mathbf{a}_3$ onto the plane of $\hat{\mathbf{a}}_\perp - \hat{\mathbf{a}}_=$ is half of $2\mathbf{a}_\perp$ and $\mathbf{a}_=$, respectively. Thus, the array marked with $m = 32$ and $n = 8$ in Fig. 7c consists of 16 and 4 big cells along $\hat{\mathbf{a}}_\parallel$ and $\hat{\mathbf{a}}_\perp$-directions, respectively. Along with $\mathbf{a}_\parallel$, only one cell layer was used with periodic boundary conditions. To get Fig. 7d-g, the 'Frequency Domain' solver was used with an input frequency of $0.49(2\pi c/\omega a)$, marked in Fig. 7a and b in the main text.

**Simulation details.** All photonic structure simulations were performed using COMSOL Multiphysics®. To input periodicity, the Floquet periodic boundary condition was imposed on all periodic boundaries. All band structures were obtained using the 'Eigenfrequency' solver.

# Data availability

The datasets generated during this study are available from the corresponding author on reasonable request.

# Code availability

The custom codes used in this study are available from the corresponding author on reasonable request.

# Acknowledgement

This work was supported by Samsung Research Funding & Incubation Center for Future Technology of Samsung Electronics (Project Number SRFC-MA1801-04). We also acknowledge the support of the European Regional Development Fund (ERDF) via the Welsh Government (80762-CU145 (East)). H.P. acknowledges the research professor fellowship, supported from by a Korea University grant.

# Author contributions

H.P. and S.L. conceived this topic. H.P. discovered the change of Weyl points' positions with $p$. H.P. also predicted and simulated the existence of pseudomagnetic fields, asymmetric localization, and the evasion behavior of the photonic wave. H.P. and S.S.O. simulated the surface potential with the Weyl equation. S.S.O. reviewed all theoretical formulations and simulations. S.L. supervised all the work. All authors contributed to a discussion of the data and the manuscript.

**Competing interests**

The authors declare that they have no competing interests.

**Additional information**

Correspondence and requests for materials should be addressed to S.S.O. and S.L.

Supplementary Information for

## "Surface potential-driven surface states in 3D topological photonic crystals"

Haedong Park[1,2], Sang Soon Oh[2]*, and Seungwoo Lee[1,3,4,5]*

[1]KU-KIST Graduate School of Converging Science and Technology, Korea University, Seoul 02841, Republic of Korea

[2]School of Physics and Astronomy, Cardiff University, Cardiff CF24 3AA, United Kingdom

[3]Department of Biomicrosystem Technology, Korea University, Seoul 02841, Republic of Korea

[4]Department of Integrative Energy Engineering and KU Photonics Center, Korea University, Seoul, 02841 Republic of Korea

[5]Center for Opto-Electronic Materials and Devices, Post-Silicon Semiconductor Institute, Korea Institute of Science and Technology (KIST), Seoul 02792, Republic of Korea

*Email: ohs2@cardiff.ac.uk, seungwoo@korea.ac.kr

# Contents

# 1. Photonic Landau levels and photonic wave localization

Applying the design in Fig. 4a in the main text to the array of 48 primitive cells generates Landau levels and Landau plateau near $\omega_0$ (Fig. S1). The perturbation strength $p$ linearly increases from $n=0$ to $n=48$. We display all these Landau spectrums along with the $\Gamma N'$- (Fig. S1a-c) and $\Gamma H'$-directions (Fig. S1d-f) ($\Gamma N'$ and $\Gamma H'$ are marked in the inset in Fig. S1d). The width of the Landau plateau is a scale of the magnetic lengths, $l_N^{-1} \sim \sqrt{B^N}$ and $l_H^{-1} \sim \sqrt{B^H}$, and marked in Fig. S1a and d, respectively. We show the results only on positive $\Gamma N'$- and $\Gamma H'$-directions because all eigenfrequencies and eigenstates are even at the $\Gamma$-point, due to time-reversal symmetry. We investigate the eigenstates along the zeroth Landau level (red curves in Fig. S1a and d) to see surface states localized on the boundary. The process of collecting eigenstates is as follows: first, for a given $\mathbf{k}$, we gather the average normal values of magnetic field eigenstates in each unit cell. These values are normalized with the maximum among these. We then collect all these sets for the $\mathbf{k}$ along the $\Gamma N'$- and $\Gamma H'$-axes, as shown in Fig. S1b and e, respectively. Localization aspects vary significantly by the $N_0$ or $H_0$. For $|\mathbf{k}| < |\Gamma N_0|$ and $|\mathbf{k}| < |\Gamma H_0|$, the eigenstates are localized on the lower boundary (which corresponds to the leftmost boundary in Fig. 4a in the main text). The confinement length on the lower boundary is a scale of the magnetic length. Around $N_0$ and $H_0$, eigenstates vary dramatically with increasing $\mathbf{k}$. For $|\mathbf{k}| > |\Gamma N_0|$ and $|\mathbf{k}| > |\Gamma H_0|$, we observe the eigenstates localized on the upper boundary (which corresponds to the rightmost boundary in Fig. 4a in the main text). These results show that $N_0$ and $H_0$, the Weyl points with $p = p_0$, are the phase transition points. We carried out the same computations with several values of $p_s$. The zeroth Landau level flattens over a larger momentum interval as the pseudomagnetic field increases (Fig. S1c and f).

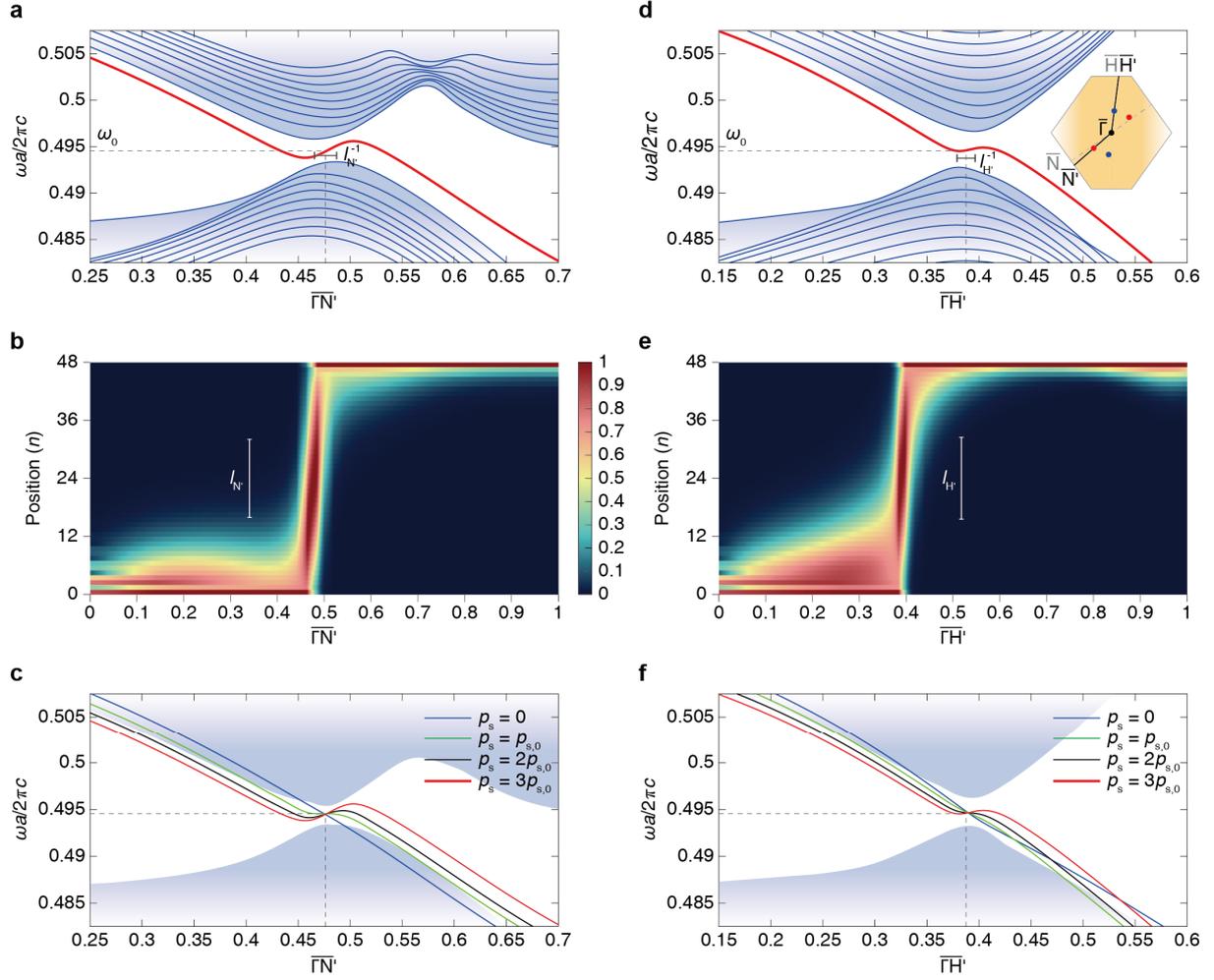

**Fig. S1. Photonic Landau levels and eigenstates in a pseudomagnetic field by a spatial gradient of** $p$. **a** and **d**, Photonic band structures along $\overline{\Gamma N'}$- (**a**) and $\overline{\Gamma H'}$-directions (**d**), which exhibit Landau plateaus at around $N_0$ and $H_0$, respectively. Here, $p_s = 3p_{s,0}$ is used where $p_{s,0} = 2.9463 \times 10^{-4} a^{-1}$. **b** and **e**, Normalized eigenstates along the zeroth Landau levels (red curves in **a** and **d**, respectively) to see surface states localized on the boundary. **c** and **f**, Zeroth Landau levels according to several magnitudes of pseudomagnetic fields. The $\overline{\Gamma N'}$ and $\overline{\Gamma H'}$-directions are marked in the inset of **d**.

## 2. Surface states by Weyl equation

Based on equation (4) in the main text, we give detailed derivation of the formula exhibiting surface states. For an orthogonal coordinate system by $x_1$, $x_2$, and $x_3$, we consider a system periodic along with $x_1$- and $x_2$-directions, and finite along $x_3$-direction, as mentioned in the main text. Thus, between the surface boundaries parallel to $x_1$- and $x_2$-directions, there are $N$ cells along $x_3$-direction. One boundary's coordinate is considered as $[x_1, x_2, 0]$, and the other is $[x_1, x_2, Na_3]$, where $a_3 = |\mathbf{a}_3|$ is the $x_3$-directional lattice constant. A point $\mathbf{x} = [x_1, x_2, x_3]$ in $n$th grid from the first boundary is discretized as $[x_1, x_2, na_3]$. The traveling wave solution $\psi(\mathbf{x}, t) = \psi_0 e^{i(\mathbf{k} \cdot \mathbf{x} - \omega t)}$ can be written as

$$\psi(\mathbf{x}, t) = u(x_3) e^{i(k_1 x_1 + k_2 x_2 - \omega t)} = u_n e^{i(k_1 x_1 + k_2 x_2 - \omega t)}, \tag{S1}$$

where $u_n = u(x_3 = na_3)$. Substituting this solution in the equation (4) in the main text gives

$$\{(k_1 - k_{w,1})\sigma_1 + (k_2 - k_{w,2})\sigma_2 - k_{w,3}\sigma_3\}u_n - i\sigma_3 \frac{\partial u_n}{\partial x_3} + V_s \sigma_3 u_n = \omega u_n, \tag{S2}$$

where $\mathbf{k}_w = [k_{w,1}, k_{w,2}, k_{w,3}]$ is the Weyl point's location, which can be decomposed as $\mathbf{k}_w = \mathbf{k}_{w,0} + \mathbf{A}_w^n$. Thus, $\mathbf{k}_w$ depends on the grid index $n$. The scalar coefficient of the surface potential $V_s$ also varies with $n$. For $1 < n < N$, $\partial u_n / \partial x_3$ is numerically converted as

$$\frac{\partial u_n}{\partial x_3} = \frac{u_{n+1} - u_{n-1}}{2a_3}. \tag{S3}$$

Thus, equation (S2) becomes

$$Du_{n-1} + C_n u_n - Du_{n+1} + V_s \sigma_3 u_n = \omega u_n, \tag{S4}$$

where

$$C_n = (k_1 - k_{w,1})\sigma_1 + (k_2 - k_{w,2})\sigma_2 - k_{w,3}\sigma_3 \tag{S5}$$

and

$$D = \frac{i\sigma_3}{2a_3}. \tag{S6}$$

Here, $C$ has a subscript $n$ due to $\mathbf{k}_w$ varies with $n$. Around the boundaries i.e., $n = 1$ and $n = N$, we use following numerical differentials:

$$\frac{\partial u_1}{\partial x_3} = \frac{-3u_1 + 4u_2 - u_3}{2a_3} \tag{S7}$$

and

$$\frac{\partial u_N}{\partial x_3} = \frac{u_{N-2} - 4u_{N-1} + 3u_N}{2a_3}. \tag{S8}$$

By substituting equations (S7) and (S8) into (S2), we have

$$(C_1 + 3D)u_1 - 4Du_2 + Du_3 + V_s\sigma_3 u_1 = \omega u_1 \tag{S9}$$

and

$$-Du_{N-2} + 4Du_{N-1} + (C_N - 3D)u_N + V_s\sigma_3 u_N = \omega u_N, \tag{S10}$$

respectively. Equations (S4), (S9), and (S10) are summarized as follows:

$$\begin{bmatrix} C_1 + 3D & -4D & D & 0 & \cdots & 0 & 0 & 0 & 0 \\ D & C_2 & -D & 0 & \cdots & 0 & 0 & 0 & 0 \\ 0 & D & C_3 & -D & \cdots & 0 & 0 & 0 & 0 \\ \vdots & \vdots & \vdots & \vdots & \ddots & \vdots & \vdots & \vdots & \vdots \\ 0 & 0 & 0 & 0 & \cdots & D & C_{N-2} & -D & 0 \\ 0 & 0 & 0 & 0 & \cdots & 0 & D & C_{N-1} & -D \\ 0 & 0 & 0 & 0 & \cdots & 0 & -D & 4D & C_N - 3D \end{bmatrix} \begin{bmatrix} u_1 \\ u_2 \\ u_3 \\ \vdots \\ u_{N-2} \\ u_{N-1} \\ u_N \end{bmatrix}$$

$$+ \begin{bmatrix} V_s\sigma_3 & 0 & 0 & \cdots & 0 & 0 & 0 \\ 0 & V_s\sigma_3 & 0 & \cdots & 0 & 0 & 0 \\ 0 & D & V_s\sigma_3 & \cdots & 0 & 0 & 0 \\ \vdots & \vdots & \vdots & \ddots & \vdots & \vdots & \vdots \\ 0 & 0 & 0 & \cdots & V_s\sigma_3 & 0 & 0 \\ 0 & 0 & 0 & \cdots & 0 & V_s\sigma_3 & 0 \\ 0 & 0 & 0 & \cdots & 0 & 0 & V_s\sigma_3 \end{bmatrix} \begin{bmatrix} u_1 \\ u_2 \\ u_3 \\ \vdots \\ u_{N-2} \\ u_{N-1} \\ u_N \end{bmatrix} = \omega \begin{bmatrix} u_1 \\ u_2 \\ u_3 \\ \vdots \\ u_{N-2} \\ u_{N-1} \\ u_N \end{bmatrix}.$$

$$\tag{S11}$$

For a system with $N = 48$, we simply use $\mathbf{a}_i = a\hat{\mathbf{x}}_i$ and $[\mathbf{b}_1 \mathbf{b}_2 \mathbf{b}_3]^T = 2\pi[\mathbf{a}_1 \mathbf{a}_2 \mathbf{a}_3]^{-1}$. We set the constant term of the Weyl point's location $\mathbf{k}_{w,0}$ as $\mathbf{k}_{w,0} = 0.194\mathbf{b}_1 - 0.22\mathbf{b}_2 + 0.1171\mathbf{b}_3$. The varying term $\mathbf{A}_w^n$ is written as $\mathbf{A}_w^n = p_s\{(n-1)a_3 - d_0\}\mathbf{k}_s$, where $\mathbf{k}_s = [6.4867, -5.5920, 1.4216]a^{-1}$ is the Weyl point's shift direction with varying $n$, and $d_0 =$ $(N/2 - 1/2)a_3$ is the distance between the midplane and the $n = 1$ grid.

We first investigate the surface localization when the surface potential is not considered (i.e., $V_s$ is zero for all the grids). When $p_s = -p_{s,0}$ and $p_s = p_{s,0}$ were respectively used ($p_{s,0}=$ 8.8388 $\times 10^{-4} a^{-1}$), the zeroth Landau levels shown in Fig. S2a-b are identical. We collect the (normalized) eigenstates intensities along with the zeroth Landau levels to observe surface states localized on the boundaries. As shown in Fig. S2c-f, the eigenstate amplitudes are mutually symmetric.

On the contrary, when we consider the $V_s$, the mutual symmetric distributions of wave intensities for the two cases are broken. We assume that $V_s$ is non-zero only on the first/last one or two grids around the boundaries, as shown in Fig. 8 in the main text; $V_s(n = 1) = -0.1 a^{-1}$, $V_s(n = 47) = 0.05 a^{-1}$, and $V_s(n = 48) = 0.1 a^{-1}$. Although several studies[1-5] have used the Dirac delta function or a similar function as $V_s$, we impose finite values as we consider the finite and discretized system. The overall shape of $V_s = V_s(n)$ is similar to Fig. 1d in the main text. The resulting surface localization is biased towards a specific boundary, as shown in Fig. S3.

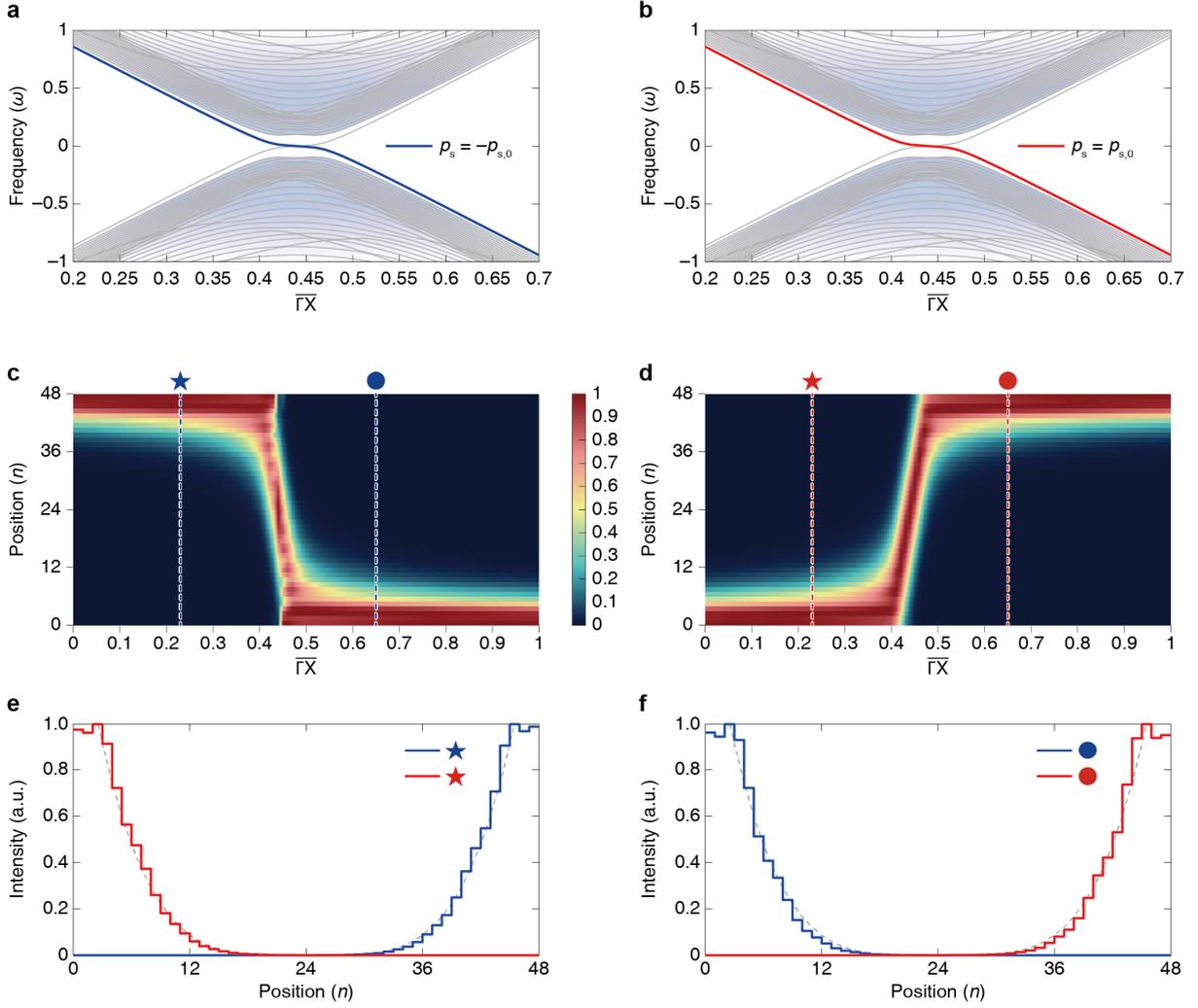

**Fig. S2. Symmetric localization of eigenstates by Weyl equation when pseudomagnetic field is applied. a**, **b**, Band structures and the zeroth Landau levels calculated by equation (S11). Each case corresponds to $p_s = -p_{s,0}$ and $p_s = p_{s,0}$, respectively, where $p_{s,0} = 8.8388 \times 10^{-4} a^{-1}$. $\overline{\Gamma X} = 0.441 \mathbf{b}_1 - 0.5 \mathbf{b}_2$ passes the projected Weyl point's location. **c**, **d**, Normalized eigenstates along the zeroth Landau levels shown in **a** and **b** to see surface states localized on the boundaries. **e**, **f**, Comparisons of the localized wave intensities along the vertical lines marked in **c** and **d**. Data with the same symbols are overlapped in the same plot. Gray dotted lines are the symmetric curves with respect to $n = 24$ for comparisons of the red and blue plots.

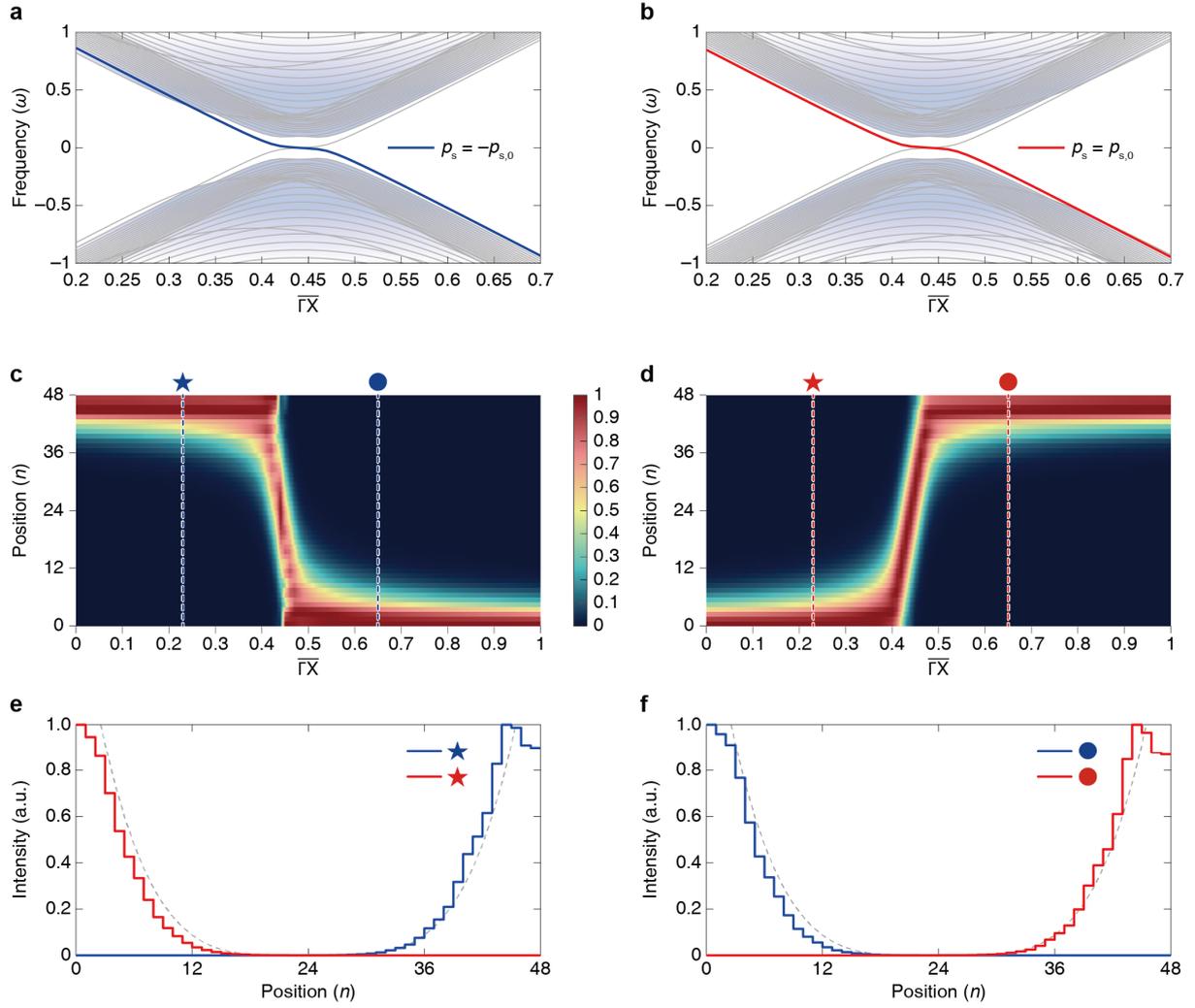

**Fig. S3. Asymmetric localization of eigenstates by Weyl equation when pseudomagnetic field and surface potential are applied.** The explanations for Fig. S2 can be used again for all the panels. The main point in here is that the wave intensities are biased towards the $n = 0$ boundary due to the application of the surface potential.

# 3. Berry phase of Weyl points at equifrequency

We introduced a double gyroid (DG) photonic crystal in Fig. 3 in the main text. The DG exhibits four point-degeneracies marked as $N_0$ and $H_0$ in Fig. 3d in the main text. To see if these are Weyl points, we calculate the Berry phase using the Wilson Loop method[6-10]. The Berry phases of the lower and upper bands connected to point $N_0$ exhibit decreasing and increasing by $2\pi$, respectively (see Fig. S4a). Therefore, its Chern number is $-1$. Likewise, the lower and upper bands connected to point $H_0$ reveals increasing and decreasing by $2\pi$, respectively (see Fig. S4b). Therefore, its Chern number is $+1$. All these are nonzero Chern numbers. Thus, these points are Weyl points, and all surface states in this study are topologically nontrivial by the Weyl points.

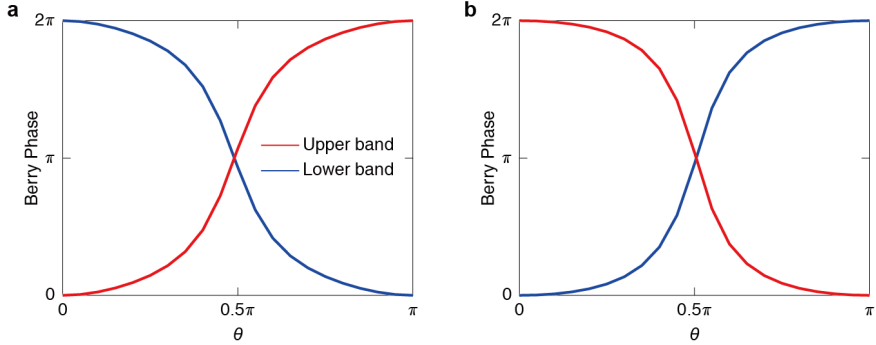

**Fig. S4. Berry phase around the equifrequency Weyl points $N_0$ (a) and $H_0$ (b) shown in Fig. 1e in the main text.** The Chern numbers of $N_0$ and $H_0$ are $-1$ and $+1$, respectively.

S10

## 4. Derivation of pseudomagnetic field in DG array

Here, we use $\mathbf{a}_1 = a/2\,[-1,1,1]$, $\mathbf{a}_2 = a/2\,[1,-1,1]$, and $\mathbf{a}_3 = a/2\,[1,1,-1]$. The normal vector of the perfect electric conductor (PEC) boundary at $n = 0$ shown in Fig. 4a in the main text is $\mathbf{a}_\perp = \mathbf{a}_2 + 0.5(\mathbf{a}_1 + \mathbf{a}_3)$. We assume that this plane passes the origin. Therefore, the distance between a point $\mathbf{x} = [x_1, x_2, x_3]$ and this plane is $d(\mathbf{x}) = \hat{\mathbf{a}}_\perp \cdot (\mathbf{x} - \mathbf{0})$ or $d(\mathbf{x}) = (x_1 + x_3)/\sqrt{2}$. The perturbation strength $p$, plotted in the inset of Fig. 2c in the main text, is given by

$$p(\mathbf{x}) = p_s\{d(\mathbf{x}) - d_0\} + p_0 \tag{S12}$$

where $p_s$ is a proportional constant, and $d_0$ is a distance between the planes at $n = 0$ and $n = N/2$, marked in the inset of Fig. 4a in the main text. The constant $p_0$ is a central value of $p(\mathbf{x})$ (see the inset of Fig. 4a in the main text) or a perturbation strength used for the default photonic crystal to generate Fig. 3a-c in the main text. If the number of unit cells between these boundaries is $N$, and the perturbation strengths at these boundaries has been determined, the proportional constant $p_s$ is calculated as follows:

$$p_s = \frac{p(n\mathbf{a}_\perp) - p(\mathbf{0})}{d(n\mathbf{a}_\perp) - d(\mathbf{0})} \tag{S13}$$

In other words, this quantity reflects the design of the DG array, such as the number of unit cells and the gradient of the perturbation strength.

As shown in Fig. 3d in the main text, Weyl points are confined to (001)-plane, irrespective of $p$. The normal direction of this plane coincides with the normal direction of the plane spanned by $\mathbf{a}_1$ and $\mathbf{a}_2$ (see Fig. S5). If we transform the coordinate system by

$$\mathbf{x}' = \begin{bmatrix} \cos\frac{\pi}{4} & -\sin\frac{\pi}{4} & 0 \\ 0 & 0 & 1 \\ -\sin\frac{\pi}{4} & -\cos\frac{\pi}{4} & 0 \end{bmatrix} \mathbf{x} \tag{S14}$$

or

$$\mathbf{x} = \begin{bmatrix} \cos\left(-\frac{\pi}{4}\right) & 0 & \sin\left(-\frac{\pi}{4}\right) \\ \sin\left(-\frac{\pi}{4}\right) & 0 & -\cos\left(-\frac{\pi}{4}\right) \\ 0 & 1 & 0 \end{bmatrix} \mathbf{x}' \quad (S15)$$

the directions of ΓN and ΓH become equivalent to $\hat{\mathbf{x}}_1' = [1,0,0]$ and $\hat{\mathbf{x}}_2' = [0,1,0]$, respectively, so that we can easily write a Hamiltonian for this system. (The relation between the directions of ΓN, ΓH, and $\mathbf{a}_i$ are marked in Fig. S5.) The $p$ with respect to the new coordinate system is given by

$$p(\mathbf{x}') = p_s \left\{ \frac{\frac{1}{\sqrt{2}}(x_1' - x_3') + x_2'}{\sqrt{2}} - d_0 \right\} + p_0 \quad (S16)$$

The effective Hamiltonian around a Weyl point is expressed as[9,11]

$$H_{eff} = \sum_{i,j}^{3} v_{ij}(k_i - k_w)\sigma_j \quad (S17)$$

where $v_{ij}$ is the anisotropic velocity tensor, $\mathbf{k}$ is the wave vector, and $\sigma_j$ is the Pauli matrices. $\mathbf{k}_w$ is $\mathbf{k}_w^H$ or $\mathbf{k}_w^N$. Here, the superscripts N or H of all terms are omitted. If the Weyl points behave like Fig. 3d in the main text, $\mathbf{k}_w$ can be decomposed into a constant term, $\mathbf{k}_{w,0}$, and a varying term, $\hat{\mathbf{k}}_s \delta k = \mathbf{A}$, i.e., $\mathbf{k}_w = \mathbf{k}_{w,0} + \hat{\mathbf{k}}_s \delta k = \mathbf{k}_{w,0} + \mathbf{A}$. For the Weyl points marked in Fig. 3d in the main text, the vector potential and pseudomagnetic field for the new coordinate system are respectively given by

$$\mathbf{A}' = \left[k_{s,1}\{p(\mathbf{x}') - p_0\}, k_{s,2}\{p(\mathbf{x}') - p_0\}, 0\right]$$
$$= \frac{p_s}{\sqrt{2}} \left\{ \frac{1}{\sqrt{2}}(x_1' - x_3') + x_2' \right\} [k_{s,1}, k_{s,2}, 0] + C \quad (S18)$$

and

$$\mathbf{B}' = \nabla_{\mathbf{x}'} \times \mathbf{A}' = \frac{p_s}{2}[k_{s,2}, -k_{s,1}, k_{s,2} - \sqrt{2}k_{s,1}]$$

$$= \frac{p_s k_{s,1}}{\sqrt{2}} \frac{[1, -\sqrt{2}, -1]}{2} - \frac{p_s}{\sqrt{2}}\left(k_{s,2} - \frac{k_{s,1}}{\sqrt{2}}\right)\frac{[-1, 0, -1]}{\sqrt{2}} \quad \text{(S19)}$$

where $C = -p_s d_0 [k_{s,1}, k_{s,2}, 0]$ is a constant term. Proportional constants $k_{s,1}$ and $k_{s,2}$ are obtained from the traces in Fig. 3d-e in the main text. Thus, unlike $p_s$ in equation (S13), the constants $k_{s,1}$ and $k_{s,2}$ do not reflect the design of the array, but they are about the nature of a three-dimensional DG Weyl photonic crystal itself. The pseudomagnetic field with respect to the original coordinate system is written as follows:

$$\mathbf{B} = p_s \left\{ \frac{k_{s,1}}{\sqrt{2}} \frac{[1, 0, -1]}{\sqrt{2}} - \frac{1}{\sqrt{2}}\left(k_{s,2} - \frac{k_{s,1}}{\sqrt{2}}\right)[0, 1, 0]\right\} = p_s (B_= \hat{\mathbf{a}}_= + B_\| \hat{\mathbf{a}}_\|) \quad \text{(S20)}$$

where $B_=$ and $B_\|$ are respectively

$$B_= = \frac{k_{s,1}}{\sqrt{2}} \quad \text{(S21)}$$

$$B_\| = -\frac{1}{\sqrt{2}}\left(k_{s,2} - \frac{k_{s,1}}{\sqrt{2}}\right) \quad \text{(S22)}$$

and $\hat{\mathbf{a}}_= = [1, 0, -1]/\sqrt{2}$ and $\hat{\mathbf{a}}_\| = [0, 1, 0]$ are the unit vectors of $\mathbf{a}_= = -\mathbf{a}_1 + \mathbf{a}_3$ and $\mathbf{a}_\| = \mathbf{a}_1 + \mathbf{a}_3$, respectively, placed on the boundary (see Fig. 4a in the main text). The number of cells $N$ and the difference of $p(\mathbf{x})$ at both boundaries are reflected only in $p_s$, as mentioned in equation (S13). Therefore, the direction of the pseudomagnetic field $\mathbf{B}$ is always fixed, regardless of the system scale and the gradient of $p(\mathbf{x})$.

The values of each variable in the above derivations are summarized in **Table S1**. All these values are about the pseudomagnetic fields derived by Weyl points around $N_0$ and $H_0$ for $N = 48$. Both pseudomagnetic fields at the antipodes of $N_0$ and $H_0$ are $-\mathbf{B}^N$ and $-\mathbf{B}^H$, respectively, due to the signs of $\mathbf{k}_s^N$ and $\mathbf{k}_s^H$ are flipped.

**Table S1. Information of pseudomagnetic fields derived by Weyl points around $N_0$ and $H_0$ for 48 cells.** All these are calculated by $p_s = 8.8388 \times 10^{-4} a^{-1}$. The magnetic length is given by $l^{-1} \sim \sqrt{B}$.

$\mathbf{k}_s^N = \begin{bmatrix} k_{s,1}^N & k_{s,2}^N \end{bmatrix} = [7.9083 \quad -3.5815]a^{-1}$ (in $\mathbf{x}'$-coordinate)

$\mathbf{B}^N = p_s B_=^N \hat{\mathbf{a}}_= + p_s B_\parallel^N \hat{\mathbf{a}}_\parallel$

$\quad = 4.9427 \times 10^{-3} a^{-2} \hat{\mathbf{a}}_= + 5.7334 \times 10^{-3} a^{-2} \hat{\mathbf{a}}_\parallel$

$\quad = [3.495 \quad 5.7334 \quad -3.495] \times 10^{-3} a^{-2}$

$\quad = [3.9541 \quad 6.4866 \quad -3.9541] p_s a^{-1}$ (in $\mathbf{x}$-coordinate)

$B^N = 7.5698 \times 10^{-3} a^{-2} = 8.5643 p_s a^{-1}$

$l_N = 11.4936 a$

$l_N^{-1} = 0.087005 a^{-1}$

$\mathbf{k}_s^H = \begin{bmatrix} k_{s,1}^H & k_{s,2}^H \end{bmatrix} = [-1.888 \quad 9.52]a^{-1}$ (in $\mathbf{x}'$-coordinate)

$\mathbf{B}^H = p_s B_=^H \hat{\mathbf{a}}_= + p_s B_\parallel^H \hat{\mathbf{a}}_\parallel$

$\quad = -1.18 \times 10^{-3} a^{-2} \hat{\mathbf{a}}_= - 6.7844 \times 10^{-3} a^{-2} \hat{\mathbf{a}}_\parallel$

$\quad = [-0.83438 \quad -6.7844 \quad 0.83438] \times 10^{-3} a^{-2}$

$\quad = [-0.94399 \quad -7.6756 \quad 0.94399] p_s a^{-1}$ (in $\mathbf{x}$-coordinate)

$B^H = 6.8862 \times 10^{-3} a^{-2} = 7.7909 p_s a^{-1}$

$l_H = 12.0506 a$

$l_H^{-1} = 0.082983 a^{-1}$

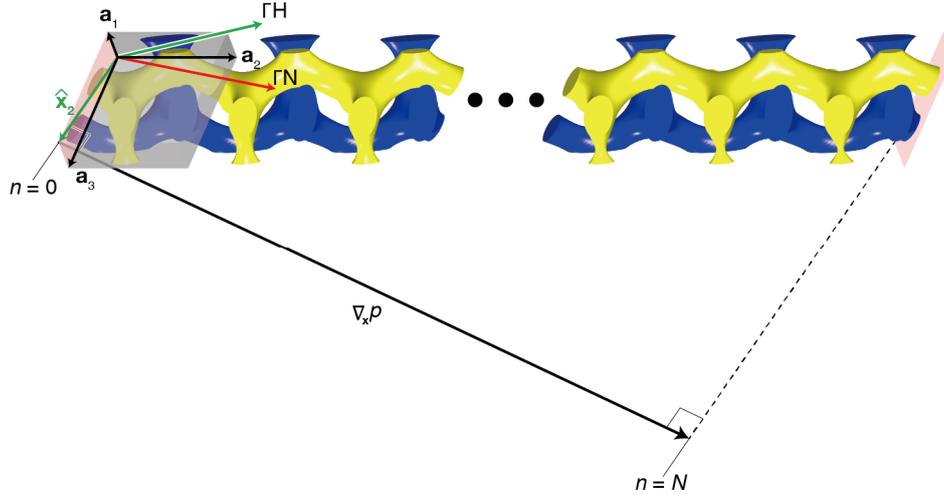

**Fig. S5. DG primitive cells array that was input to get Fig. 4-6 in the main text.** The array consists of $N$ cells with spatially linearly changing $p$ along $\hat{\mathbf{a}}_\perp$-direction. Several directions significantly dealt with in this study are also marked.

**References**

1. Potter, A. C., Kimchi, I. & Vishwanath, A. Quantum oscillations from surface Fermi arcs in Weyl and Dirac semimetals. *Nature Communications* **5**, 5161 (2014).
2. Bhowmick, S. & Shenoy, V. B. Weber-Fechner type nonlinear behavior in zigzag edge graphene nanoribbons. *Physical Review B* **82**, 155448 (2010).
3. Shtanko, O. & Levitov, L. Robustness and universality of surface states in Dirac materials. *Proceedings of the National Academy of Sciences* **115**, 5908-5913 (2018).
4. Dou, Z. et al. Imaging Bulk and Edge Transport near the Dirac Point in Graphene Moiré Superlattices. *Nano Letters* **18**, 2530-2537 (2018).
5. Dongre, N. K. & Roychowdhury, K. Effects of surface potentials on Goos-Haenchen and Imbert-Fedorov shifts in Weyl semimetals. *arXiv preprint arXiv:2106.04573* (2021).
6. Park, H. & Lee, S. Double Gyroids for Frequency-Isolated Weyl Points in the Visible Regime and Interference Lithographic Design. *ACS Photonics* **7**, 1577-1585 (2020).
7. Yu, R., Qi, X. L., Bernevig, A., Fang, Z. & Dai, X. Equivalent expression of Z2 topological invariant for band insulators using the non-Abelian Berry connection. *Physical Review B* **84**, 075119 (2011).
8. Wang, Q., Xiao, M., Liu, H., Zhu, S. & Chan, C. T. Optical Interface States Protected by Synthetic Weyl Points. *Physical Review X* **7**, 031032 (2017).
9. Soluyanov, A. A. et al. Type-II Weyl semimetals. *Nature* **527**, 495 (2015).
10. Yang, Z. et al. Weyl points in a magnetic tetrahedral photonic crystal. *Opt. Express* **25**, 15772-15777 (2017).
11. Jia, H. et al. Observation of chiral zero mode in inhomogeneous three-dimensional Weyl metamaterials. *Science* **363**, 148-151 (2019).